\documentclass[%
 reprint,
 superscriptaddress,
 amsmath,amssymb,
 aps,
]{revtex4-2}

\usepackage{graphicx}% Include figure files
\usepackage{dcolumn}% Align table columns on decimal point
\usepackage{xcolor}

\usepackage{hyperref}
\hypersetup{
    colorlinks=true,
    linkcolor=blue,
    filecolor=black,      
    urlcolor=black,
    citecolor= black
}

\usepackage{bm}% bold math
\begin{document}

%\preprint{APS/123-QED}

\title{High fidelity two-qubit gates on fluxoniums using a tunable coupler}% Force line breaks with \\

\author{Ilya N. Moskalenko}
\email[Corresponding author: ]{in.moskalenko@misis.ru}
\affiliation{National University of Science and Technology ``MISIS'', 119049 Moscow, Russia}
\affiliation{Russian Quantum Center, 143025 Skolkovo, Moscow, Russia}

\author{Ilya A. Simakov}
\affiliation{National University of Science and Technology ``MISIS'', 119049 Moscow, Russia}
\affiliation{Russian Quantum Center, 143025 Skolkovo, Moscow, Russia}
\affiliation{Moscow Institute of Physics and Technology, 141701 Dolgoprudny, Russia}

\author{Nikolay N. Abramov}
\affiliation{National University of Science and Technology ``MISIS'', 119049 Moscow, Russia}
\affiliation{Russian Quantum Center, 143025 Skolkovo, Moscow, Russia}
\author{Alexander A. Grigorev}
\affiliation{National University of Science and Technology ``MISIS'', 119049 Moscow, Russia}

\author{Dmitry~O.~Moskalev}
\affiliation{Dukhov Research Institute of Automatics (VNIIA),  Moscow 127055, Russia}
\affiliation{FMN Laboratory, Bauman Moscow State Technical University, Moscow 105005, Russia}
\author{Anastasiya~A.~Pishchimova}
\affiliation{Dukhov Research Institute of Automatics (VNIIA),  Moscow 127055, Russia}
\affiliation{FMN Laboratory, Bauman Moscow State Technical University, Moscow 105005, Russia}
\author{Nikita~S.~Smirnov}
\affiliation{Dukhov Research Institute of Automatics (VNIIA),  Moscow 127055, Russia}
\affiliation{FMN Laboratory, Bauman Moscow State Technical University, Moscow 105005, Russia}
\author{Evgeniy~V.~Zikiy}
\affiliation{Dukhov Research Institute of Automatics (VNIIA),  Moscow 127055, Russia}
\affiliation{FMN Laboratory, Bauman Moscow State Technical University, Moscow 105005, Russia}
\author{Ilya~A.~Rodionov}
\affiliation{Dukhov Research Institute of Automatics (VNIIA),  Moscow 127055, Russia}
\affiliation{FMN Laboratory, Bauman Moscow State Technical University, Moscow 105005, Russia}
\author{Ilya S. Besedin}
\affiliation{National University of Science and Technology ``MISIS'', 119049 Moscow, Russia}
\affiliation{Russian Quantum Center, 143025 Skolkovo, Moscow, Russia}

%\author{Alexey V. Ustinov}
%\affiliation{National University of Science and Technology "MISIS", 119049 Moscow, Russia}
%\affiliation{Russian Quantum Center, 143025 Skolkovo, Moscow, Russia}
%\affiliation{Physikalisches Institut, Karlsruhe Institute of Technology, Karlsruhe, Germany}

\date{\today}% It is always \today, today,

\begin{abstract}
Superconducting fluxonium qubits provide a promising alternative to transmons on the path toward large-scale superconductor-based quantum computing due to their better coherence and larger anharmonicity. A major challenge for multi-qubit fluxonium devices is the experimental demonstration of a scalable crosstalk-free multi-qubit architecture with high fidelity single-qubit and two-qubit gates, single-shot readout and state initialization. 
Here, we present a two-qubit fluxonium-based quantum processor with a tunable coupler element following our theoretical proposal \cite{Moskalenko2021}. 
We experimentally demonstrate fSim-type and controlled-Z gates with $99.55\%$ and $99.23\%$ fidelities, respectively. The residual ZZ interaction is suppressed down to the few kHz level. Using a galvanically coupled flux control line, we implement high fidelity single-qubit gates and ground state initialization with a single arbitrary waveform generator channel per qubit.

% \begin{description}
% Secondary publications and information retrieval purposes.
% You may use the \texttt{description} environment to structure your abstract;
% use the optional argument of the \verb+\item+ command to give the category of each item. 
% \end{description}
\end{abstract}

\maketitle

%\tableofcontents

\section{\label{sec:introduction}INTRODUCTION }

Superconducting qubits have become one of the most successful platforms for quantum computing during the past decade. One of the pillars of this success was the development of the transmon qubit \cite{Koch2007}. The typical transmon-based toolkit consists of a coplanar waveguide (CPW) resonator for dispersive readout \cite{Blais2004} and capacitive coupling that facilitates two-qubit gates \cite{Yamamoto2003}. 
The key limitation of transmon-based quantum computing is dielectric loss, which limits the qubit coherence time. Incremental progress in material science and fabrication technology over the years has enabled an increase in coherence times from few microseconds \cite{Houck2008} to hundreds of microseconds \cite{Place2021, Wang2022}. Despite this remarkable progress, dielectric loss is still a major issue on the route to large scale quantum computing with superconducting qubits. Another fundamental issue with transmon qubits is their low relative anharmonicity, which leads to longer gate times and ultimately lower gate fidelities.
Nevertheless, transmon qubits have been hugely successful for the development of noisy intermediate scale quantum information processing devices\cite{Arute2019, Wu2021}. Recent implementations of two-qubit gates on transmons demonstrate two-qubit gate fidelities around 99.5\% \cite{Foxen2020, Sheldon2016, Abrams2020, Negirneac2020, Noguchi2020, Leek2019}. Another major  issue for large-scale devices is crosstalk suppression. Among transmon qubits, one of the most critical types of crosstalk is static ZZ interaction. Recently, tunable couplers have been widely used as a tool to mitigate ZZ crosstalk in a scalable and effective way \cite{Yan2018, Sung2021, Mundada2019, Xu2020, Collodo2020}.

Development towards large-scale quantum computing motivates the search of multi-qubit architectures with the better gate fidelities and simpler control system. 

\begin{figure*}
    \includegraphics[width=2\columnwidth]{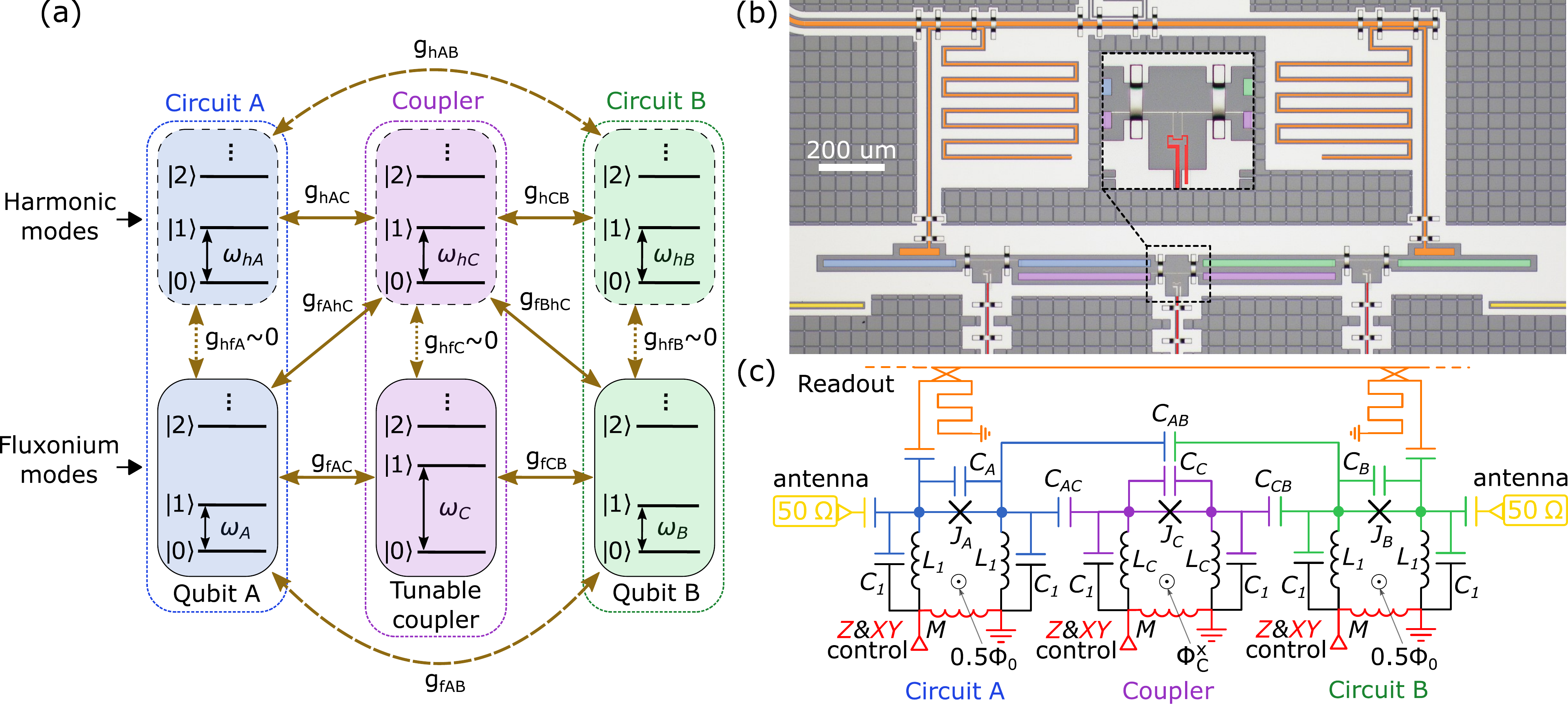}
    \caption{(a) Schematic diagram of an interacting three-body system. (b) Experimental realization of three capacitively coupled fluxonium-like qubits fabricated on silicon substrate. (c) Circuit schematic. In (b), false colors (blue, purple,  green, orange, yellow and red) are used to indicate the corresponding circuit components in (d). $50~\Omega$ terminators are installed at 10~mK stage of the dilution refrigerator and used for qubits initialization (see Appendix \ref{sec:Qubits_initialization} for details). }
    \label{fig:Device}
\end{figure*}

Superconducting fluxonium qubits \cite{Manucharyan2009, Nguyen2019, Pop2014, Zhang2021} could provide a possible alternative to the transmon path due to their lower transition frequencies and high coherence properties. Microwave-activated CZ gates \cite{Ficheux2021} have already been demonstrated on the fluxonium platform in a 3D cavity. Another example of high-fidelity two-qubit gates \cite{Bao2021} has been shown in a more scalable approach with two qubits coupled to separate readout resonators, integrated on a single chip. Development of a tunable capacitive coupling architecture could provide means for Fluxonium-based NISQ devices.

In this paper, we experimentally demonstrate high-fidelity two-qubit gates between low-frequency fluxonium qubits using a tunable capacitive coupler setup shown in Fig.~\ref{fig:Device}a, which was proposed earlier in our theoretical work \cite{Moskalenko2021}. Both qubits and coupler are modified fluxonium circuits with additional harmonic modes. The computational qubits are operated at the flux degeneracy point where the qubit transition frequency lies in the 600-750~MHz range. Low frequencies lead to operation in a relatively hot environment. To initialize the qubit in the ground state, before each measurement, we employ a reset flux pulse that rapidly tunes the qubit frequency to several GHz. At this higher frequency the thermal equilibrium corresponds to negligible excited state occupation. The circuit was fabricated by using the aluminum shadow evaporation process and allows to implement resource-efficient qubit control (qubit biasing and excitation are performed via the same control line) with a single arbitrary waveform generator (AWG) channel per qubit. Each qubit is driven directly by the AWG device, without any need of IQ mixers and high-frequency local oscillators.
We demonstrate suppression of static ZZ interaction by using our tunable coupler design and achieve average single-qubit gate fidelities above $99.96 \%$ using gaussian excitation pulses with 13.3~ns duration. The realised 60~ns-long two-qubit parametric-resonance fSim-type gate \cite{Sete2021} demonstrates $99.55 \pm 0.04\%$ fidelity in cross-entropy benchmarking (XEB) \cite{Arute2019}. We construct a 183.3~ns-long pulse sequence that implements the CZ gate with $99.23 \pm 0.04\%$ fidelity. The CZ pulse sequence consists of two fSim gates and five single-qubit gates and uses a spin-echo technique to cancel the effect of conditional phase accumulation during the fSim gate.

\section{\label{sec:level1_Device}DEVICE DESCRIPTION }

The two-qubit quantum processor consists of three capacitively coupled fluxonium circuits with additional harmonic modes, see~Fig.\ref{fig:Device}a. An optical image of the processor chip is shown in~Fig.\ref{fig:Device}b.
The lowest two energy levels of the fluxonium modes of the left the and right circuits A and B are used as data qubits. The middle circuit C is a two-mode tunable coupler, and is designed to always remain in the ground state. This scheme allows to realize a fSim-type gates \cite{Arute2019, Foxen2020}. 

The electric schematic of the two-qubit system is shown in fig.~\ref{fig:Device}c. The Hamiltonian of the system can be written as:
\begin{equation}
\label{eq:full Hamiltonian}
	\hat{H}_\textnormal{full} = \sum_{i=f_A,f_B,f_C}^{} \hat{H}_{i} + \sum_{j=h_A,h_B,h_C}^{} \hat{H}_{j} +
	\sum_{ij}^{} \hat{g}_{ij}.
\end{equation}
The fluxonium ($f_C$) and harmonic ($h_C$) modes of the coupler at the center mediate interaction between two distant fluxonium qubits ($f_A$ and $f_B$). Transition frequencies of qubit A ($Q_A$) $\omega_A/2\pi$ and qubit B ($Q_B$) $\omega_B/2\pi$ at their flux degeneracy points are equal to 688.224~MHz and 664.763~MHz respectively, and can be tuned up to 3~GHz by applying an external magnetic flux through their loops. The measured coherence times of the qubits at the flux degeneracy point are $T_{1,A}= 87~\mathrm{\mu s}$, $T_{1,B}= 86~\mathrm{\mu s}$, $T_{2,A}= 51~\mathrm{\mu s}$, $T_{2,B}= 46~\mathrm{\mu s}$, $T^E_{2,A}= 107~\mathrm{\mu s}$, $T^E_{2,B}= 93~\mathrm{\mu s}$. More details about the device are provided in Appendix  \ref{sec:Device_parameters}.
The harmonic mode frequency of the coupler ($h_C$) is $\omega_{hC}/2\pi=2.0~\mathrm{GHz}$. The frequency of the coupler fluxonium mode ($\omega_{CF}$) is tunable, similar to the qubits. Qubits and coupler frequencies are controlled via individual galvanically coupled fast flux bias lines. 

As described in \cite{Moskalenko2021}, after eliminating the coupler degrees of freedom and harmonic modes ($h_A$, $h_B$), our two-qubit processor obeys the effective low-energy Hamiltonian:
\begin{equation}
	\hat{H}_\textnormal{eff}/\hbar = -\frac{1}{2}\omega_A\sigma^\textnormal{z}_A - \frac{1}{2}\omega_B\sigma^\textnormal{z}_B + g_\textnormal{xx}\sigma^\textnormal{x}_A\sigma^\textnormal{x}_B + \frac{1}{4}\zeta_\textnormal{zz}\sigma^\textnormal{z}_A\sigma^\textnormal{z}_B.
	\label{eq6}
\end{equation}

\section{\label{sec:Single_qubit_gates}SINGLE-QUBIT GATES VIA DIRECT RF SYNTHESIS}

We realize single-qubit gates and individual frequency control via a single flux bias line to each qubit. This approach has been recently demonstrated for transmon qubits \cite{Arute2019, Manenti2021}. We use a Zurich Instruments HDAWG8 Arbitrary Waveform Generator (AWG), which has a 2.4~GHz sampling rate. In contrast to transmon qubits, where the frequency range of frequency control pulses is far apart from the frequency range required for qubit excitation, for fluxonium qubits both DC signals and qubit frequency signals can be generated from a single AWG channel. 

A common issue for flux control lines in superconducting qubits is attenuation and filtering: to access the full range of tunability of a superconducting qubit, one needs to be able to induce at least half of a magnetic flux quantum in the qubit loop. If the mutual inductance between the control line and the qubit is small, a large current amplitude in the control line is required. If the mutual inductance is large, the control line becomes a significant decay channel for the qubit excited state. The design value of the mutual inductance in our fluxonium circuits is 12~pH, which corresponds to a qubit lifetime of $1$~ms and bias current of $83~\mathrm{\mu A}$ to shift the qubit into the flux degeneracy point. At the same time, the qubit drive $\Omega$ rate per unit AC current is $\Omega/I \approx 320~\mathrm{MHz/\mu A}$.

\begin{figure}
    \includegraphics[width=1\columnwidth]{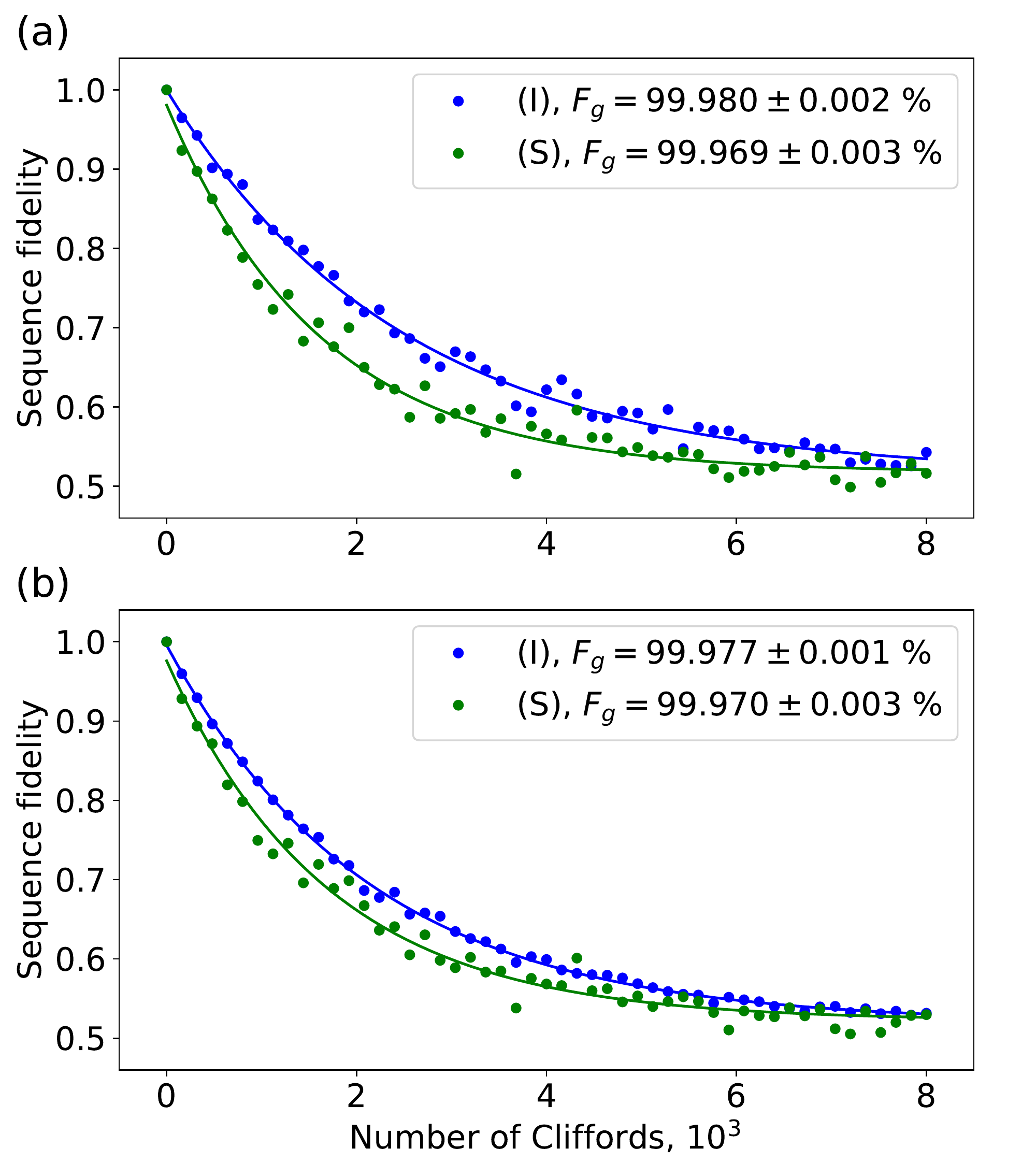}
    \caption{Experimental results of randomized benchmarking for single-qubit gates on qubit A and qubit B. (a) Measurement data of single-qubit RB for qubit A. (b) Measurement data of single-qubit RB for qubit B. "I" denotes the isolated application of single-qubit Cliffords and isolated qubit readout. "S" denotes the simultaneous application of single-qubit Cliffords and joint qubit readout. Error estimates for the fidelity are obatined from the least square fit errors. The data are averaged over 20 random sequences for each sequence length.}
    \label{fig:1Q_benchmarking}
\end{figure}

We use the small excitation amplitude regime for single-qubit gate operation $\Omega \ll \omega$. In this regime, the rotating wave approximation is valid, and counter-rotating terms in the driven Hamiltonian can be neglected. However, the excitation pulse amplitude is still sufficient to significantly shift the qubit from the flux degeneracy point, resulting in phase errors. We use two different approaches to overcome these phase errors. 
Our first approach is based on the Derivative Removal by Adiabatic Gate (DRAG) technique \cite{Motzoi2009}. The excitation signal is given by
\begin{equation}
\begin{aligned}
    \Omega(t) = & \epsilon_x(t)\sin \omega t + \alpha\dot{\epsilon}_x(t)\cos \omega t, \\
    \epsilon_x(t) = & A\left( e^{-\frac{t^2}{2\sigma^2}} - e^{-\frac{T_\mathrm{p}^2}{2\sigma^2}} \right).
\end{aligned}
\end{equation}

We pick $T_\mathrm{p}=13.3~\mathrm{ns}$, which corresponds to the minimum waveform length of the AWG, and $\sigma=3.3~\mathrm{ns}$. Even though the nature of phase errors in fluxonium qubits is very different from phase errors in transmon qubits, adding an extra signal into the orthogonal quadrature can remove the effect of the frequency shift to first order. 
Using a sequence of calibration measurement, we find the values of $A$ corresponding to $\pi/2$ and $\pi$ pulses. Using a procedure based on the amplified phase error technique \cite{Lucero2010}, we find $\alpha$.

The other approach to solving the phase error problem is virtual Z-gates \cite{McKay2017}. Here we only use one quadrature for the $\pi/2$ pulse, and correct for the phase error by adding a phase increment to the AWG's numerically controlled oscillator (NCO). The phase increment is instant, and thus the Z-gate has unit fidelity. All single-qubit gates can be divided into three families: virtual Z-gates, denoted as $U_1(\varphi)$, gates consisting of one $\pi/2$-pulse and a two Z-gates $U_2(\varphi, \lambda)$, and all other single-qubit gates $U_3(\theta, \varphi, \lambda)$, which require two $\pi/2$ gates and three Z-gates.
\begin{figure*}
    \includegraphics[width=1.5\columnwidth]{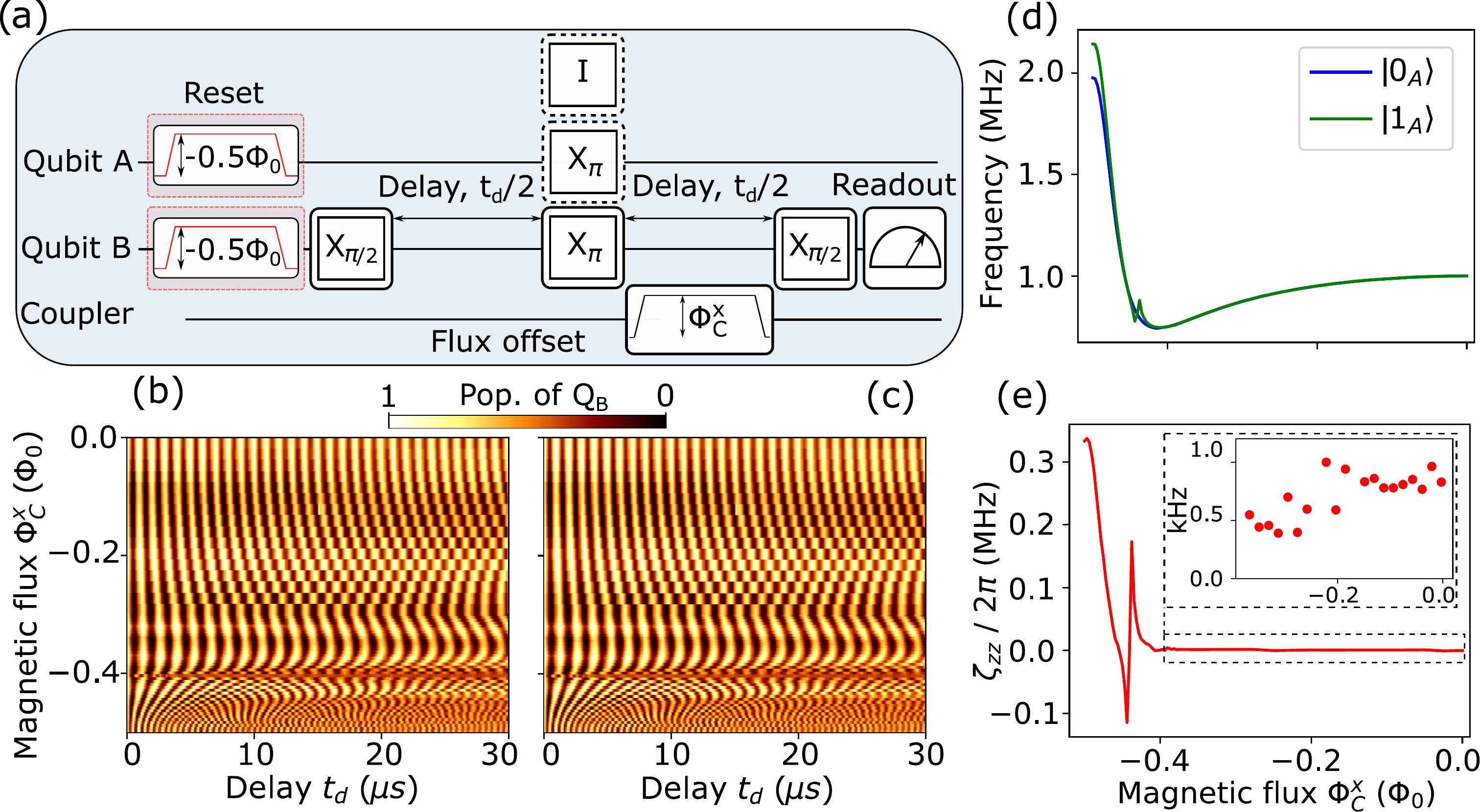}
    \caption{Measurement of ZZ interaction strength $\zeta_\mathrm{zz}$. (a) Pulse sequence of an echo-type experiment on qubit B while initializing qubit A in its ground (preparation gate $I$) or exited (preparation gate $X_{\pi}$) state. The final $\pi/2$ pulse phase is modulated proportionally to the delay. (b) Qubit B population measurement results without the $\pi$ pulse applied to qubit A. (c) Qubit B population measurement results with the $\pi$ pulse applied to qubit A. (d) Frequency of qubit B echo-type oscillations as a function of the coupler magnetic flux $\Phi^x_C$. The blue curve corresponds to no excitation of qubit A and the green curve to excitation of qubit A in parallel to the $\pi$ pulse applied to qubit B. (e) ZZ interaction strength $\zeta_\mathrm{zz}$ as a function of $\Phi^x_C$. The inset shows $\zeta_\mathrm{zz}$ values near the zero flux point in the coupler.}
    \label{fig:ZZ_measurements}
\end{figure*}
We use the DRAG approach to demonstrate the high gate fidelities accessible to our two-qubit device. For all other measurements in this paper measurements we use the Z-gate approach, because it can be efficiently implemented in the native instructions of the AWG sequencer in a manner that is compatible with our reset pulse (Appendix~\ref{sec:Qubits_initialization}). 

Measurement of single-qubit gate fidelity is carried out by Clifford-based randomized benchmarking (RB) \cite{Magesan2012a,Corcoles2013,Barends2014}. The Clifford group is generated from an identity pulse, a $\pi/2$-pulse, a $\pi$-pulse, and a virtual-Z gate that corresponds to a $\pi/2$ rotation. Among different decompositions of Clifford group gates, we pick the decomposition that has minimal duration, yielding an average of $5/6$ 13.3~ns-long pulses per gate. The results of single-qubit RB are shown in Fig.~\ref{fig:1Q_benchmarking}. We repeat the experiment for qubit A, qubit B, and qubit A and B simultaneously.
The average Clifford group gate fidelities drop from 99.980~$\%$ to 99.969~$\%$ when switching from seprate qubit benchmarking to simultaneus benchmarking for qubit A, and from 99.977~$\%$ to 99.970~$\%$ for qubit B. We attribute this degradation to residual $\sigma_x\sigma_x$ coupling between the qubits when the coupling is turned off.

\section{\label{sec:TWO_qubit_gates}TWO-QUBIT GATES USING A TUNABLE COUPLER}

Our two-qubit device allows to implement  universal two-qubit gates from the fSim family \cite{Arute2019, Foxen2020}, which describes the set of excitation number-preserving quantum logic operations on two qubits up to single-qubit phase rotations.
Its matrix representation in the $|00\rangle$, $|01\rangle$, $|10\rangle$, $|11\rangle$ basis is given by:
\begin{equation}
    \operatorname{fSim}(\theta, \varphi)=\left(\begin{array}{cccc}
    1 & 0 & 0 & 0 \\
    0 & \cos \theta & -i \sin \theta & 0 \\
    0 & -i \sin \theta & \cos \theta & 0 \\
    0 & 0 & 0 & e^{-i \varphi}
    \end{array}\right),
    \label{c}
\end{equation}
where $\theta$ is the swap angle, and $\varphi$ is the conditional phase.

We use the notation $|Q_A, Q_B\rangle$ to represent the eigenstates of the system in the idling configuration, when $C_F$ is placed at it's maximum frequency ($\Phi^x_C=0$) such that the effective qubit-qubit coupling is close to zero. Both qubits are parked at their minimal frequencies %(flux degeneracy points $\Phi^x_{Q1,2}=0.5\Phi_0$) 
while idling and during single-qubit gates.

To experimentally characterize the dependence of $\zeta_{\mathrm{zz}}$ on the coupler flux bias we use a echo-like pulse sequence shown in Fig.~\ref{fig:ZZ_measurements}a. The echo experiment is performed on qubit B. Parallel to the time reversal $\pi$ pulse applied to qubit B we can excite qubit A. After the reversal pulse, we apply a flux bias pulse to the coupler. Additionally, we introduce an additional phase shift to the final $\pi/2$ pulse that is proportional to the delay time, so that even without a flux pulse we obtain 1~MHz oscillations. Both the coupler flux pulse and the the excitation pulse applied to qubit A lead to frequency shifts of qubit B, changing the oscillation frequency. The measurement results are shown in Fig.~\ref{fig:ZZ_measurements}b-d. The doubled frequency difference between the oscillation for excited and deexcited qubit A yields the ZZ interaction strength shown in Fig.~\ref{fig:ZZ_measurements}e. 

The measurement results confirm that in our processor the static ZZ interaction $\zeta_{\mathrm{zz}}$ is nearly eliminated (less than 1 kHz) over a wide range of the magnetic fluxes in the coupler loop. 

\begin{figure}
    \includegraphics[width=1\columnwidth]{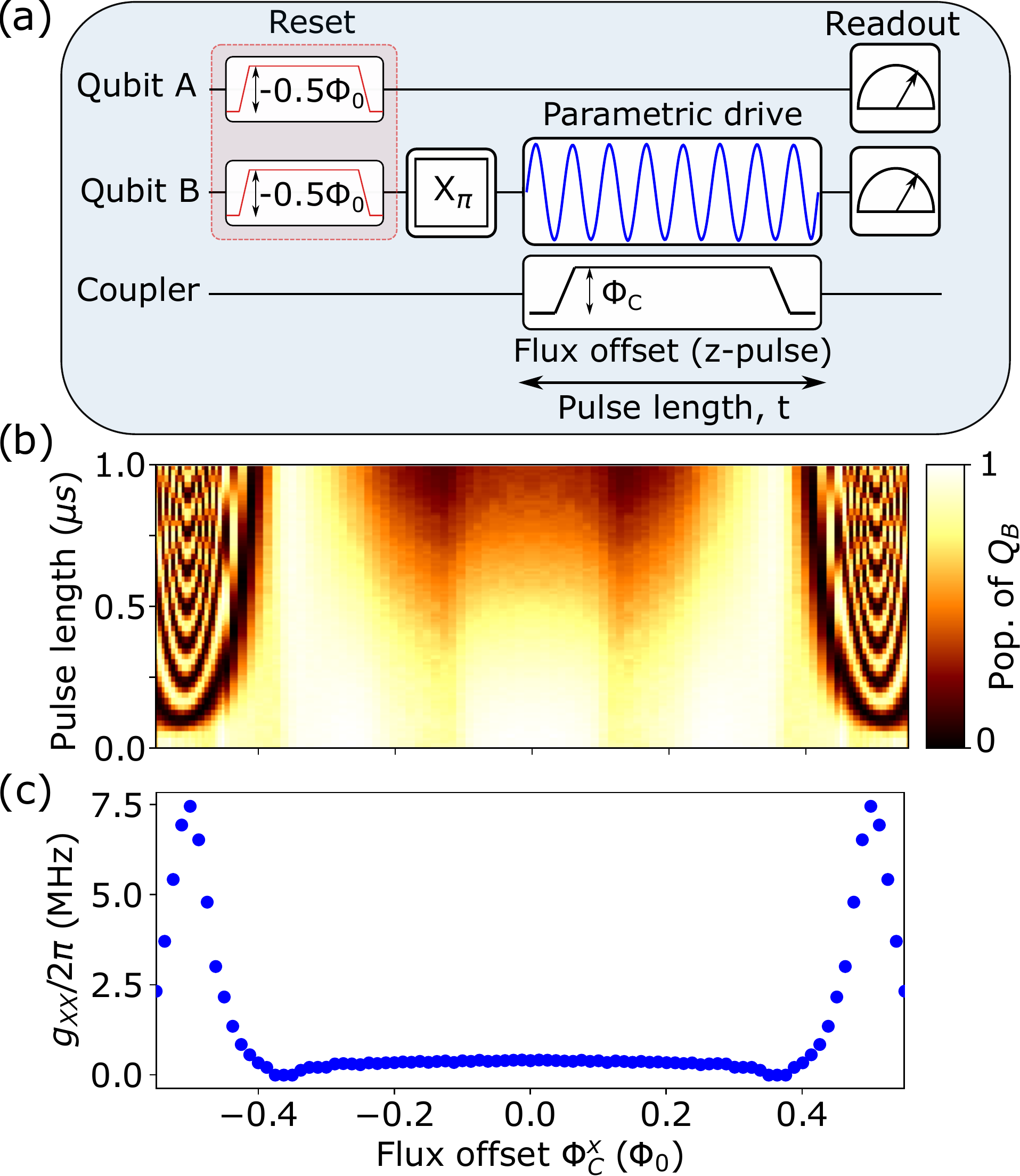}
    \caption{Effective XX coupling strength as a function of coupler flux. (a) Pulse sequence to measure qubit-qubit coupling under modulation (qubit B tuned in resonance with qubit A) as a function of the coupler flux bias. (b) Experimental data for the energy exchange between $|10\rangle$ and $|01\rangle$, as a function of the magnetic flux in the coupler $\Phi^x_C$. (c) The effective qubit-qubit coupling $g_{\mathrm{xx}}$ versus $\Phi^x_C$.}
    \label{fig:iSWAP_RABI_data}
\end{figure}

We aim at implementing a $\sqrt{\mathrm{iSWAP}}$-like gate by diabatically inducing vacuum Rabi oscillations between the $|10\rangle$ and $|01\rangle$ states. The oscillation rate is controlled by adjusting $\omega_C$, which effectively tunes the coupling strength $g_{\mathrm{xx}}$ [Fig.\ref{fig:iSWAP_RABI_data}c]. 

The frequency difference between the qubits in their working points is approximately $\Delta=23~\mathrm{MHz}$, while the maximum effective coupling strength is around $g_{\mathrm{xx}}^{\mathrm{max}}\approx8~\mathrm{MHz}$. Ramping up the effective coupling strength between the qubits alone is insufficient to realize an entangling gate: the vacuum Rabi oscillation amplitude, estimated by the formula $4g_{\mathrm{xx}}^2/(4g_{\mathrm{xx}}^2+\Delta^2)$, is below the 0.5 value required for an $\sqrt{\mathrm{iSWAP}}$ gate. 
The straightforward solution to the frequency detuning problem is to apply a flux pulse to qubit B's control line simultaneously with the flux pulse to the coupler. However, this approach undermines the very idea of using a fluxonium qubit for higher coherence due to the large dephasing rates of superconducting qubits outside of flux sweet spots.

To overcome this obstacle, we employ the parametric frequency shift technique introduced for transmon qubits in \cite{Sete2021}.  Instead of applying a rectangular-shaped flux pulse to the qubit, the frequency is shifted with an AC signal. This significantly reduces the impact of low-frequency flux noise on qubit coherence. The amplitude of this modulation is chosen so that the effective frequency of qubit B becomes equal to qubit A's frequency. The calibration procedure for the parametric frequency shift pulse is described in Appendix \ref{sec:Parametric_drive}.

\subsection{\label{sec:iSWAP_calibration}Gate calibration}

We first characterize the effective qubit-qubit coupling strength $g_{\mathrm{xx}}$ by measuring the energy exchange between $|10\rangle$ and $|01\rangle$ states as a function of the coupler flux bias. To measure the energy exchange, we prepare the $|01\rangle$ state by applying a $\pi$-pulse to qubit B followed by a modulated flux pulse $\Phi_B(t)$ (frequency $\omega_p/2\pi=100~\mathrm{MHz}$, amplitude $\tilde{\Phi}=0.016~\Phi_0$) to bring it into resonance with qubit A for a variable pulse duration $t_d$ [Fig.\ref{fig:iSWAP_RABI_data}a]. During the parametric modulation we vary the effective qubit-qubit coupling strength by biasing the coupler flux with a square-shaped flux pulse with amplitude $\Phi_C^x$. The populations of the qubits are measured as a function of $t_d$ and $\Phi_C^x$ [Fig.\ref{fig:iSWAP_RABI_data}b].

To quantify the qubit-qubit coupling strength $g_{\mathrm{xx}}$ we fit the oscillations of the $|01\rangle$ state population with harmonic oscillations at every coupler flux offset. The frequencies of the fitted harmonic oscillations are shown in Fig.~\ref{fig:iSWAP_RABI_data}c.
As expected from the simulations \cite{Moskalenko2021}, the maximum value of coupling strength occurs at the flux degeneracy point of the coupler ($\Phi_C=0.5\Phi_0$) and it also stays relatively constant $g_{\mathrm{xx}}\approx 0.4~\mathrm{MHz}$ in range from $-0.3\Phi_0$ to $0.3\Phi_0$. 

\subsection{\label{sec:fSim_metrology}fSim metrology}

Here we consider an $\sqrt{\mathrm{iSWAP}}$-like gate, $\operatorname{fSim}(\theta, \varphi$) with $\theta = -\pi/4$. The gate consists of a parametric frequency shift pulse applied to qubit B, and a flux pulse to the coupler. %The coupler flux pulse shape at the AWG is described by the formula
%\begin{equation}
%    \Phi_\mathrm{C}^{\mathrm{x}}(t) = A
%    \begin{cases}
%        \frac{2 + \sqrt{2}}{4}\sin \frac{\pi}{4}\left(\frac{t}{T_\mathrm{e}}+1\right), & 0>t>T_\mathrm{e}, \\
%        1, & T_\mathrm{e}>t>T_\mathrm{p} - T_\mathrm{e}, \\
%        \frac{2 + \sqrt{2}}{4}\sin \frac{\pi}{4}\left(\frac{t-T_\mathrm{p}+T_\mathrm{e}}{T_\mathrm{e}}+2\right), & T_\mathrm{p} - T_\mathrm{e}>t>T_\mathrm{p}.
%    \end{cases}
%\end{equation}
The total gate duration is 60~ns, which corresponds to 6 periods of the parametric frequency shift signal. To avoid leakage outside the computational subspace, we replace the front and back edges of the coupler flux pulses with fragments of a sine wave. The edge times are chosen such that the gate amplitude is as close to $0.5\Phi_0$ as possible, resulting in 14~ns for both front and back edge. After the amplitude calibration procedure described in Appendix \ref{sec:fSIM_calibration} we obtain a pulse amplitude of $0.49787\Phi_0$.

To verify the fidelity of the fSim gate, we use cross-entropy benchmarking (XEB), described in detail in \cite{Arute2019}. The main feature of the method is that it allows not only to evaluate the accuracy of a target gate, but also to estimate unknown parameters in the gate unitary matrix. Apart from the $\theta$ and $\varphi$ angles explicitly identified in the $\operatorname{fSim}$ definition \eqref{c}, our gate also includes three independent single-qubit phase rotations. 

Another important issue is that the frame of the XX coupling, which generates the $\operatorname{fSim}$ gate, is distinct from the single-qubit gate frames. The relative phase of the NCOs used for single-qubit unitaries depends on time as $\left(\omega_A-\omega_B\right)t$; this phase also enters the non-diagonal matrix elements of the unitary matrix. Similarly,  virtual Z-gates implemented by phase shifts of the NCOs have no effect on the $\operatorname{fSim}$ gate. Thus, effectively we have two relevant frames describing the device: the single-qubit rotation frame, which is defined by the oscillators in waveform generators controlling the qubits, and the laboratory frame, where the qubits rotate at different frequencies. Simulations of the system evolution under both single-qubit and $\operatorname{fSim}$ gates must account for the frame changes.

The idea of the XEB method is similar to randomized benchmarking (RB) and interleaved randomized benchmarking (IRB). Unlike in RB and IRB, after executing a sequence of random gates we do not apply a cancellation gate that returns the qubit into an eigenstate of the measurement operator, but measure the system in a random superposition state. For each circuit depth ($m$), measurements are performed for a large number (100) of different random sequences.

The gate parameter estimation is performed by comparing the measured probability distributions with probability distributions simulated for sequences of ideal unitary gates. By maximizing linear cross-entropy between simulation and experiment, we obtain estimates for the gate parameters: $\theta/\pi=0.2502$, $\varphi/\pi=0.0255$.
After finding the unitary gate parameters, we estimate the average depolarization fidelity of the final state $\overline{\varepsilon_{m}}$. It can be approximated by the function $a p^m$, where $p$ is the depolarizing parameter and $a$ is fitting parameter used to account for state preparation and measurement (SPAM) errors. The average fidelity of the executed gates is given by 
\begin{equation}
    F = p + (1-p)/D,
    \label{eq:Fdep}
\end{equation}
where $D=2^n$ is the dimension of the Hilbert space ($n=2$). If a target gate is inserted after each single-qubit operation, the average fidelity of the gate is determined by formula \eqref{eq:Fdep} with $p = p_2/p_1$, where $p_2$ and $p_1$ are the depolarizing parameter corresponding to the gate sequences with and without (reference sequences) the interleaved gate.

\begin{figure}
    \includegraphics[width=1\columnwidth]{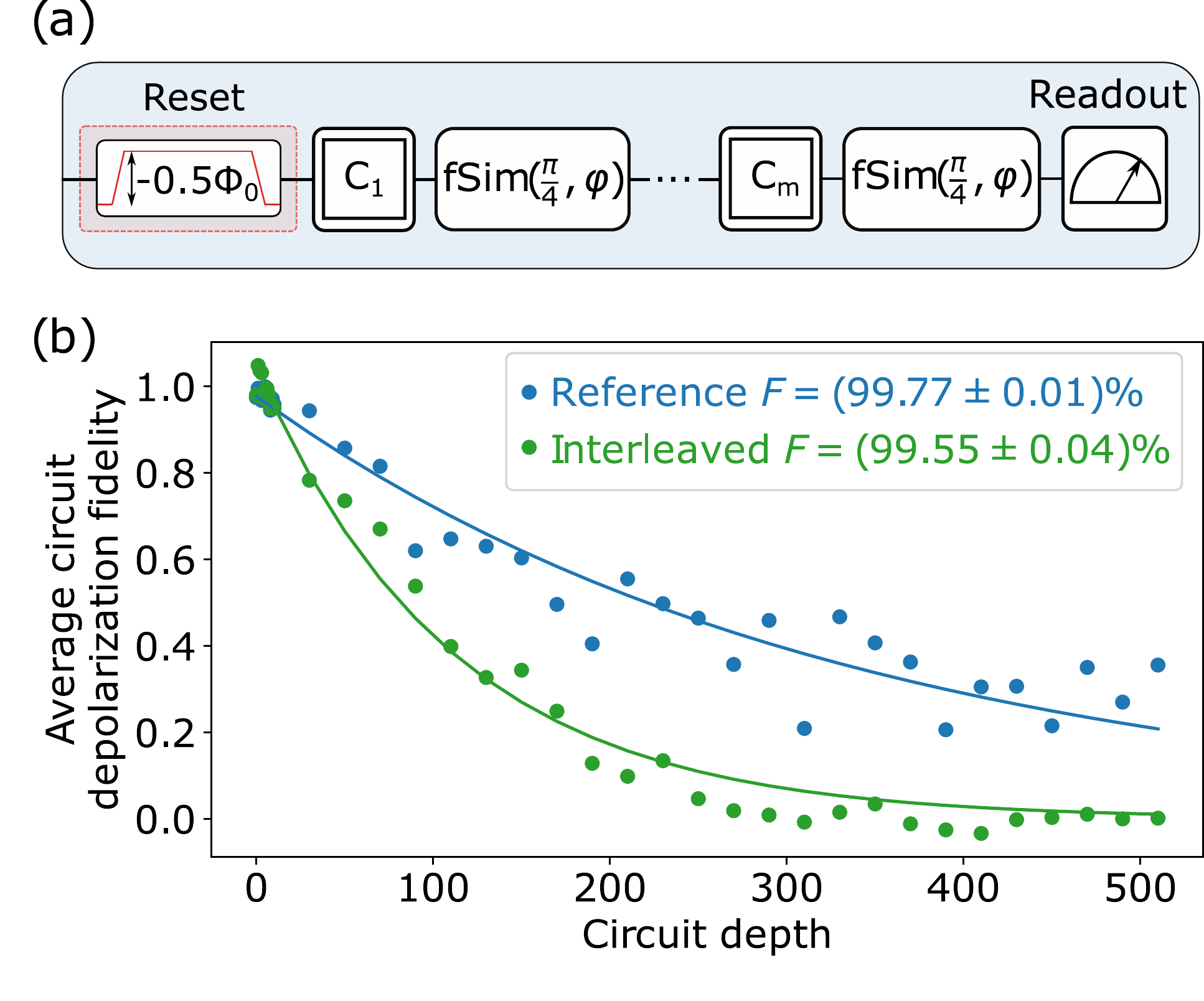}
    \caption{Cross-entropy benchmarking (XEB) of the $\sqrt{\mathrm{iSWAP}}$-like gate. (a) Pulse sequence for the XEB experiment. (b) Depolarization fidelity for the reference circuits and interleaved circuits.}
    \label{fig:fSIM_XEB_data}
\end{figure}

In Fig.~\ref{fig:fSIM_XEB_data}(b) the blue dots show the exponential decay of the depolarization fidelity $\overline{\varepsilon_{m}}$ of the reference random single-qubit Clifford-gate sequences executed simultaneously on two qubits. The green dots show similar data when an fSim gate is inserted between single-qubit operations, as shown in Fig.~\ref{fig:fSIM_XEB_data}(a). From a least-squares fit we obtain $p_1 = (99.697 \pm 0.016)\%$ and $p_2 = (99.10 \pm 0.04)\%$. The resulting fidelity of the fSim gate is $F = (99.55 \pm 0.04)\%$.

\color{black}

\subsection{\label{sec:CZ_data_metrology}CZ gate calibration and metrology}

The $\operatorname{fSim}$ gate that can be naturally implemented in our two-qubit device has a number of drawbacks. Firstly, it is the issue of two different frames for single-qubit gates and two-qubit gates that is discussed above. Secondly, an entangling Clifford-group two-qubit gate, such as iSWAP or CZ, would be more helpful to compare the performance of our device with other two-qubit gate implementations. Finally, Clifford-group two-qubit gates have the advantage of being the primitive used in many algorithms.

\begin{figure}
    \includegraphics[width=1\columnwidth]{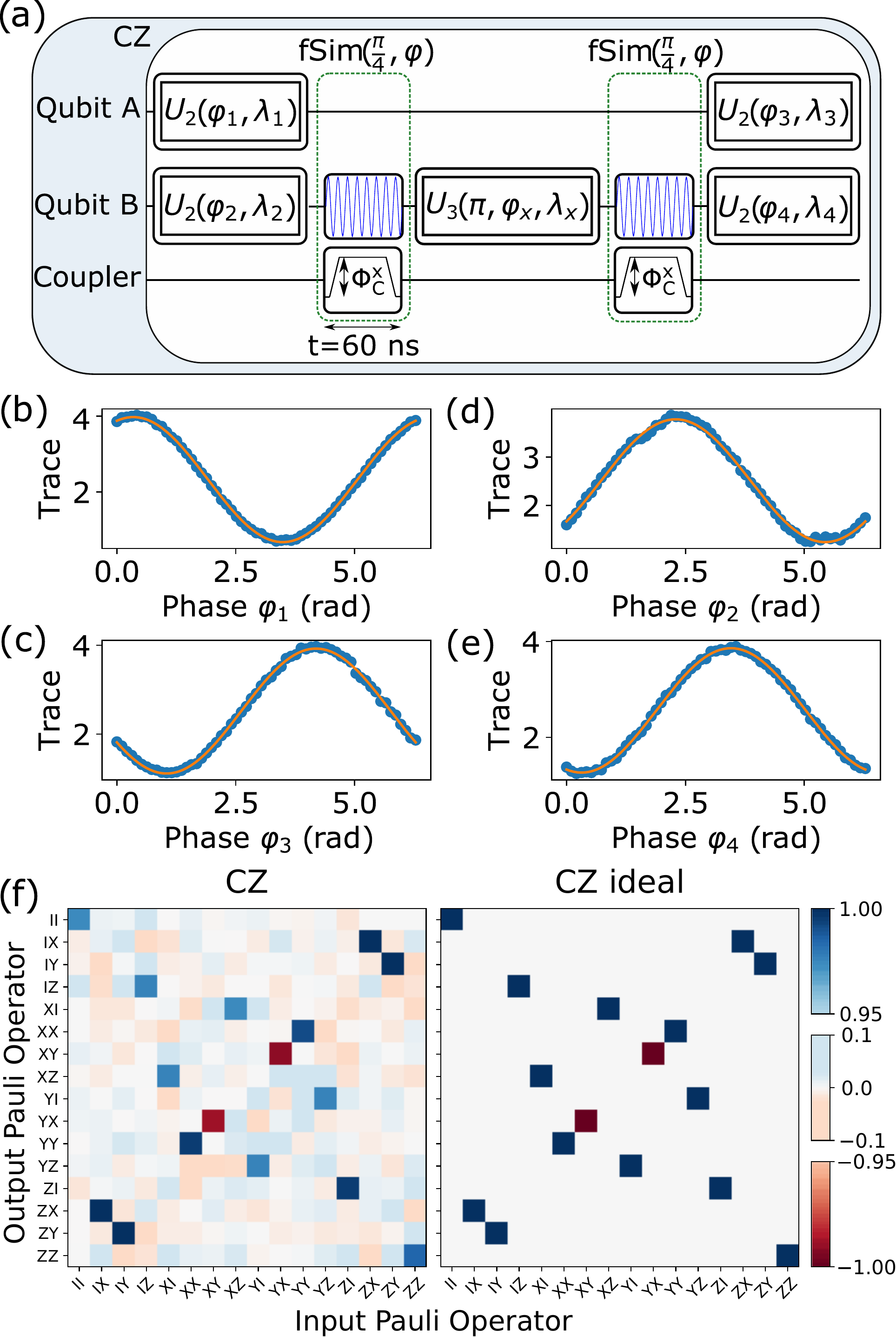}
    \caption{Controlled-Z gate calibration and quantum process tomography results. (a) Gate sequence equivalent to CZ. (b,c,d,e) Calibration data for $\varphi_1$, $\varphi_2$, $\varphi_3$, $\varphi_4$ (blue dots) and approximation with cosine function. (f) Pauli transfer matrix of the calibrated CZ gate (on the left) with fidelity $99.4\%$ obtained from quantum process tomography. Pauli transfer matrix of the ideal CZ gate (on the right) is shown for comparison.}
    \label{fig:CZ_tomography}
\end{figure}

To resolve these problems we construct a CZ gate from two $\operatorname{fSim}(\frac{\pi}{4}, \varphi)$ gates and 5 single-qubit gates, using the sequence proposed in the Supplementary information of \cite{Moskalenko2021}. The sequence is shown in terms of pulses applied to different control channels in Fig.~\ref{fig:CZ_tomography} (a).
At the core of this pulse sequence lie two identical $\operatorname{fSim}$ gates interleaved by a $\pi$-pulse, denoted as $U_3(\pi, \varphi_x, \lambda_x-\varphi_x)$. This echo-like sequence cancels out the conditional phase accumulation $\varphi$ in the $\operatorname{fSim}$ gates. The resulting gate is up to single-qubit gates equivalent to CZ. Furthermore, it can be shown that these single-qubit gates can be expressed in terms of $\pi/2$ pulses, denoted as $U_2(\varphi_i, \lambda_i-\varphi_i), i\in\{1,2,3,4\}$.

A further key premise of this sequence is that the CZ operator is diagonal in the computational basis. Thus, it remains invariant under the unitary transform connecting the single-qubit gate frame and the two-qubit gate frame. Once we construct a pulse sequence in the laboratory frame that implements CZ, this pulse sequence can be used at any point in time. Unlike the pulses used to implement standalone single-qubit gates, the single-qubit gates in this sequence should always have the same modulation phase in the laboratory frame. This ensures that relative phase between the $\operatorname{fSim}$ gate and the single-qubit gates remains the same. 

\begin{figure}
    \includegraphics[width=1\columnwidth]{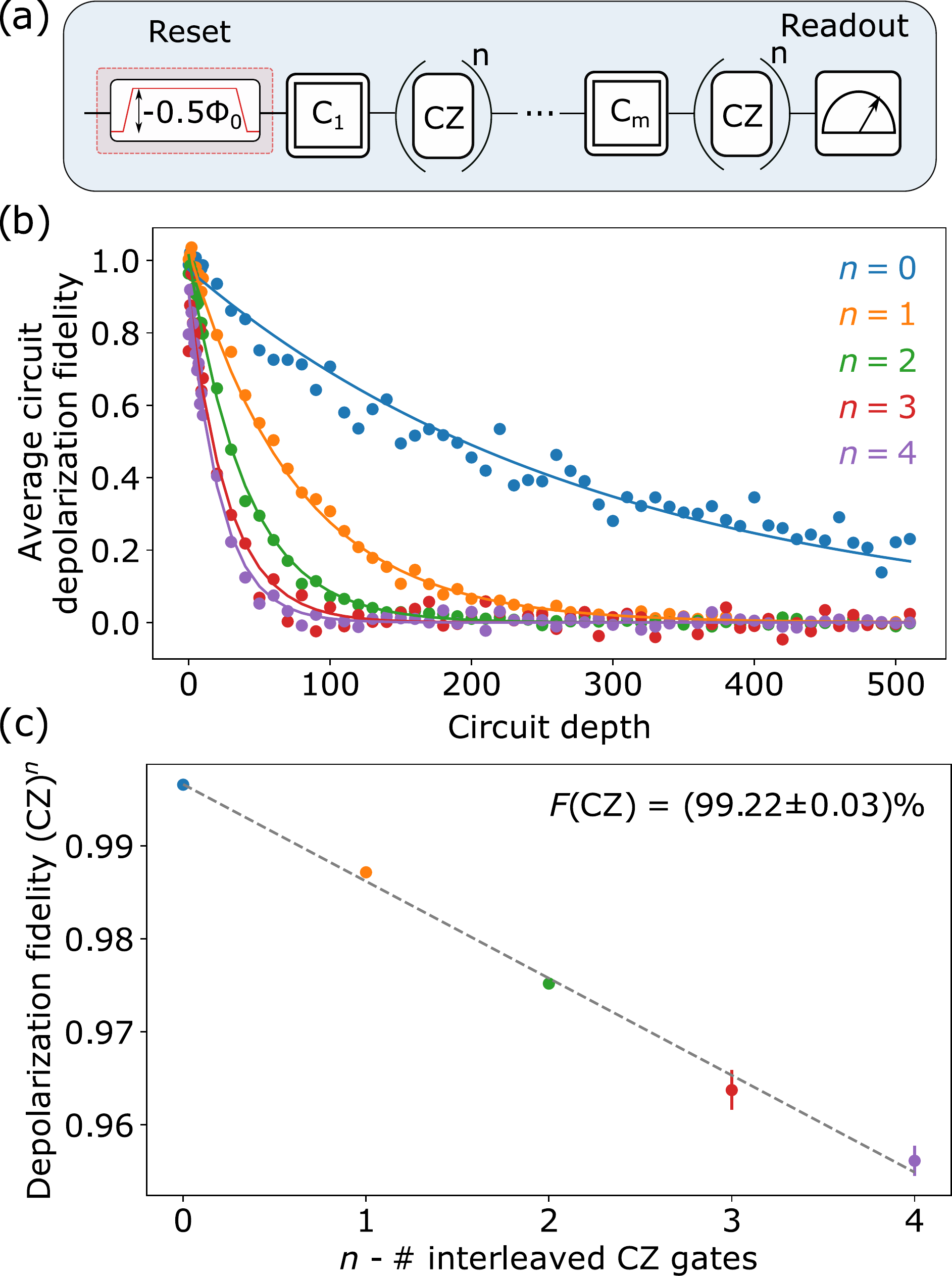}
    \caption{XEB for the CZ gate. (a) Pulse sequence for XEB experiment with the CZ gate. (b) XEB results of the CZ gate. We use a variable number of CZ gates (up to 4) (encoded in the color) between the randomly chosen Clifford gates. For each circuit depth, we average over 100 sequences. (c) Fidelity of ${CZ}^n$ versus n. Linear dependence indicates the incoherent errors.}
    \label{fig:CZ_XEB}
\end{figure}

The CZ calibration procedure starts with a known $\pi/2$ single qubit gate pulse amplitude and a known pulse sequence corresponding to $\operatorname{fSim}(\pi/4, \varphi)$. We require no prior knowledge about the relative phases of either of these pulses. Let $U$ be the unitary matrix describing the action of the pulses shown in Fig.~\ref{fig:CZ_tomography} (a).
For any values of relative phases $\lambda_1, \lambda_2, \lambda_3, \lambda_4, \lambda_x, \varphi_x$ there exists a set of values $\varphi_1, \varphi_2, \varphi_3, \varphi_4$ for which $U$ is diagonal. Furthermore, the squared absolute values of the diagonal entries of $U$ can be expressed as
\begin{equation}
\label{eq:diagonal-population}
    |u_{ii}|^2 = \prod \limits_{j=1}^4 \cos^2 \frac{\varphi_j-\varphi_j^o}{2},
\end{equation}
where $\varphi_j^o$ correspond to the sought-after values of $\varphi_j$ which correspond to a diagonal unitary $U$.

We probe the sum of the squared absolute values of the diagonal entries of $U$ by preparing each of the computational states, performing the gate sequence, and measuring the population retained in the prepared state. This resulting quantity can be interpreted as the trace of the classical transition probability matrix. Note that according to equation~\eqref{eq:diagonal-population},  all the diagonal entries of the transition probability matrix are equal, and are products of functions of $\varphi_1$, $\varphi_2$, $\varphi_3$, $\varphi_4$. Therefore, the values $\varphi_1^o$, $\varphi_2^o$, $\varphi_3^o$, and $\varphi_4^o$ can be found using only single-dimensional scans over $\varphi_1$, $\varphi_2$, $\varphi_3$, and $\varphi_4$. The measured transition probability matrix traces are shown in Fig.~\ref{fig:CZ_tomography} (b)-(e).

After obtaining a pulse sequence that corresponds to a diagonal operator that is up to single-qubit operations equivalent to the CZ gate, there are only two free parameters. These parameters are the angles of single-qubit rotations around the $Z$ axis. We determine these angles by quantum process tomography. After supplementing the pulse sequence with virtual Z-rotations to revert these rotations, we repeat process tomography. The resulting Pauli transfer matrix is shown in Fig.~\ref{fig:CZ_tomography} (f), and the gate fidelity obtained from this measurement is $99.4 \%$.

Finally, we estimate the fidelity of the CZ gate by iterative XEB. The quantum circuit of the experiment in shown in Fig~\ref{fig:CZ_XEB}(a). The experiment is performed for $n$ sequential CZ gates interleaved with random single-qubit Clifford gates. Fig~\ref{fig:CZ_XEB}(b, c) shows the depth dependence of the average depolarization fidelity. The obtained fidelity of the iterated CZ gates varies linearly with their number, which  indicates the absence of coherent errors.  From a linear fit we obtain a gate fidelity of $F = (99.22 \pm 0.03)\%$ per single gate. The errors are obtained by the standard variance of the least squares method and conventional formulas for calculating errors.

\color{black}

%From the calibration curves we defined appropriate values:
%$\varphi_1=0.3464, \varphi_2=2.2356, \varphi_3=4.2192, \varphi_4=3.4563$.

\section{CONCLUSIONS}

We have demonstrated two-qubit quantum processor based on the modified fluxonium qubits in a tunable coupler architecture and realized high-fidelity fSIM and CZ gates on the same device.
To implement two-qubit gates, we used a parametric flux modulation for bringing the qubits into a resonance with each other. As our qubits have low transition frequencies, here we also proposed and implemented unconditional reset mechanism for qubits initialization. 
The tunable coupling scheme helped us to obtain high-fidelity two-qubit operations and suppress residual ZZ-coupling rate (here less than 1 kHz), allowing for parallel high-fidelity single-qubit operations.

Taken together, this work reveals an interesting and promising approach towards fault-tolerant quantum computing with low-frequency qubits that can be good alternative and competitive to the transmon system. We believe that the low frequency of data qubits opens the possibility of using sub-gigahertz wiring and electronics for gate operations and individual qubit control, which in turn allows to reduce the complexity of control system via using a single flux bias line for each qubit.

\begin{acknowledgments}
The authors acknowledge Alexey Ustinov for helpful discussions and comments on the manuscript. Experimental part of this work was performed with the financial support from the Russian Science Foundation, Project № 21-72-30026. Devices were fabricated at the BMSTU Nanofabrication Facility (Functional Micro/Nanosystems, FMNS REC, ID 74300).
\end{acknowledgments}

\appendix

\section{\label{sec:Fab}Device design and fabrication}

\begin{figure}
    \includegraphics[width=1\columnwidth]{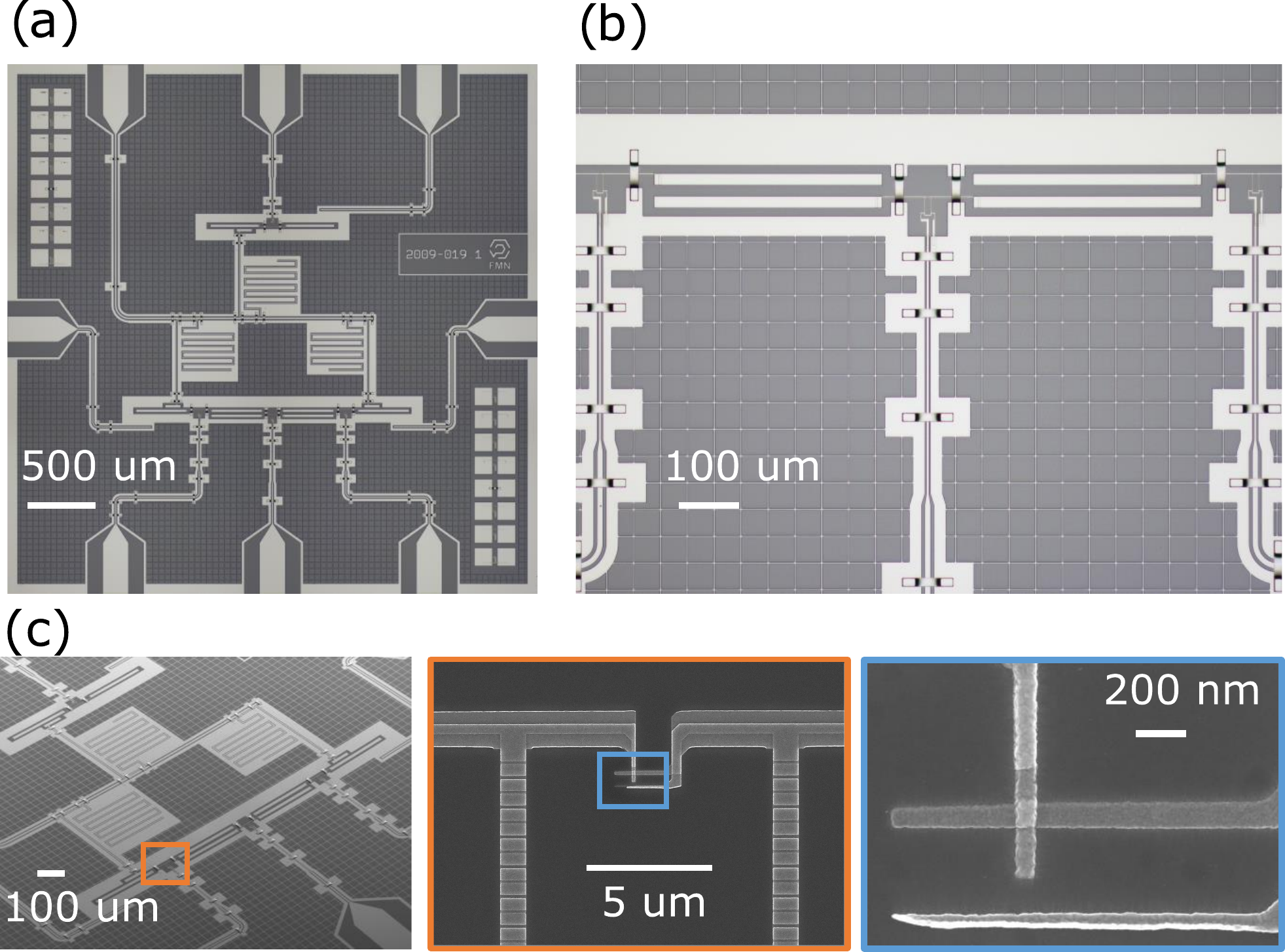}
    \caption{(a) The optical image of the chip. (b) The optical image of coupler two qubits and coupler. (c) SEM images of the fluxonium device, magnified image of the junctions array of the qubit A and the single junction.}
    \label{fig:Fabrication}
\end{figure}

The design of the investigated system consists of the tunable two-qubit quantum processor (lower part of the chip in fig.~\ref{fig:Fabrication}a), described earlier in sec. \ref{sec:level1_Device}, and a single qubit ($Q_S$) (upper part of the chip in fig.~\ref{fig:Fabrication}a), which is also implemented as a two-mode fluxonium circuit, and is used for test measurements.

The device is made in a four-step process: (I) Base Al layer patterning, (II) Josephson junction double-angle evaporation and lift-off, (III) patterning and deposition of bandages, (IV) crossovers fabrication.
 
Devices are fabricated on Topsil Global Wafers high-resistivity silicon substrate ($\rho > 10000$~Ohm · cm, $525~\mu m$ thick). Prior to the deposition the substrate is cleaned in a Piranha solution at 80°C, followed by dipping in $2\%$ hydrofluoric bath to remove the native oxide. 100 nm thick base aluminum layer is grown using e-beam evaporation in a ultra-high vacuum deposition system. 600~nm thick Dow MEGAPOSIT SPR 955-CM photoresist is then spin coated.  Qubit capacitors, resonators, wiring and ground plane are defined using a laser direct-writing lithography system (Heidelberg Instruments uPG 101), developed in AZ Developer to minimize film damage and then dry etched in BCl3/Cl2 inductively coupled plasma (Oxford PlasmaPro100). The photoresist is stripped in N-methyl-2-pyrrolidone at 80°C for 3 h and rinsed in IPA (isopropyl alcohol) with sonication.
 
The substrate is then spin coated with resist bilayer composed of 500 nm MMA (methyl methacrylate) and 300 nm PMMA (poly methyl methacrylate). The development is performed in a bath of MIBK/IPA 1:3 solution followed by rinse in IPA. Al/AlOx/Al Josephson junctions are patterned using electron beam lithography system (Raith Voyager) and aluminum electrodes are shadow-evaporated in ultra-high vacuum deposition system. 25 nm thick first Al junction electrode is oxidized at 5 mbar to form the tunnel barrier and next the 45 nm thick counter-electrode is evaporated. We then pattern and evaporate aluminum bandages using the same process as for junctions with an in-situ Ar ion milling in order to provide good electrical contact of the junction with the base layer.  Lift-off is performed in a bath of N-methyl-2-pyrrolidone with sonication at 80°C for 3 h and rinsed in a bath of IPA with sonication. 
 
Finally, aluminum free-standing crossovers are fabricated using conventional approach \cite{Chen2014a}. SPR 220 3 um photoresist is spin coated and then the sacrificial layer is patterned using a direct laser writing system. The development is performed in AZ Developer / deionized water solution (1:1) for 2 minutes in order to minimize film damaging and the resist is reflowed at 140°C. 300 nm of Al is then evaporated with an in-situ Ar ion milling to remove the native oxide. Second layer of 3 um SPR 220 is used as a protective mask and the excess metal in dry etched in inductively coupled plasma. Damaged layer of photoresist is then removed in oxygen plasma and both layers of photoresist are stripped N-methyl-2-pyrrolidone at 80°C.

\section{\label{Exp_setup} Experimental setup}

\begin{figure*}
    \includegraphics[width=1.5\columnwidth]{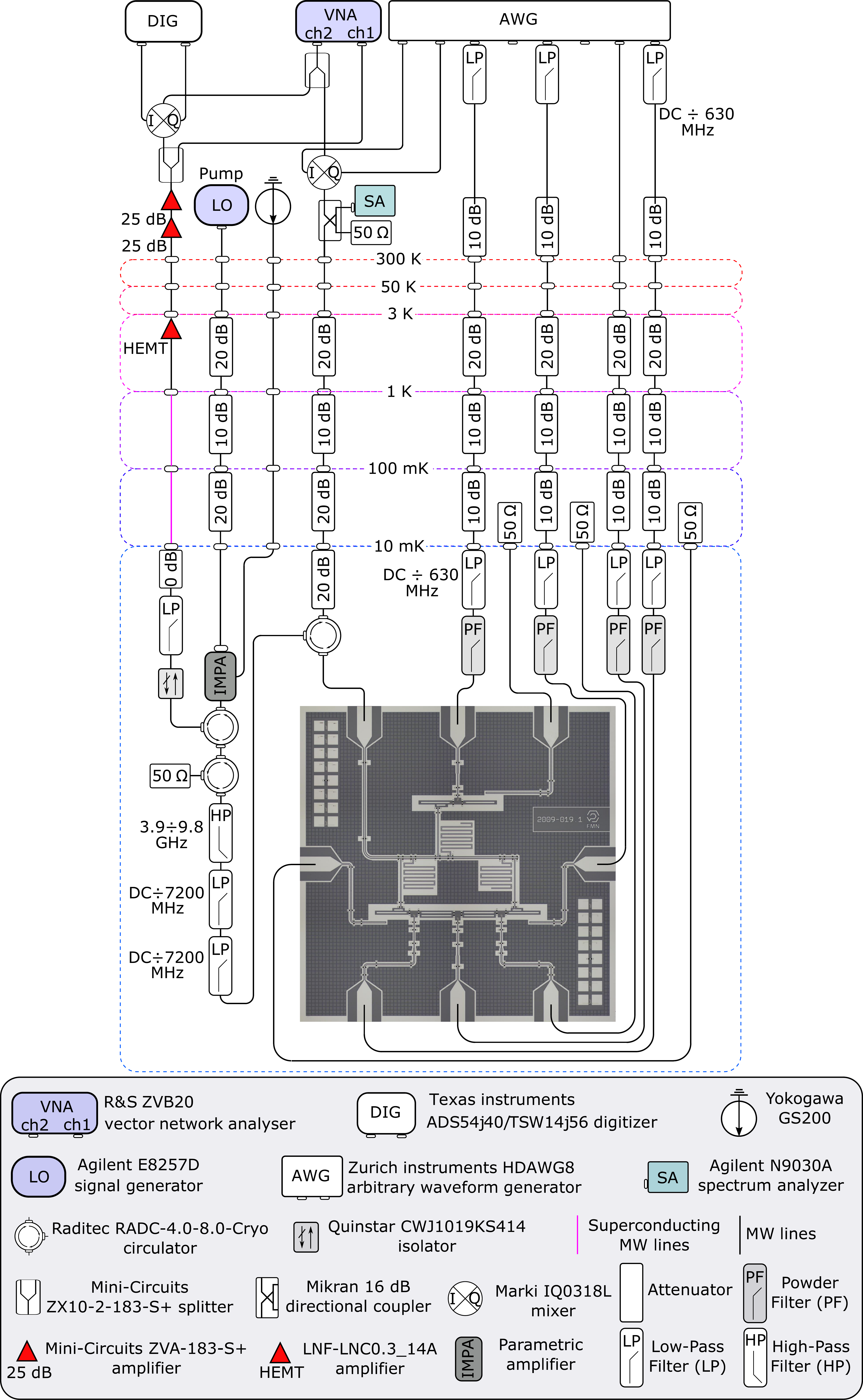}
    \caption{Schematic diagram of the experimental setup. Single-qubit control pulses modulated with qubit frequencies, flux offsets, and two-qubit gates applied to the coupler are generated directly using one analog output port of a the arbitrary waveform generator.}
    \label{fig:experimental_setup}
\end{figure*}

The experiments are performed in a BlueFors LD-250 dilution refrigerator with a base temperature of 10 mK, as shown in Fig.\ref{fig:experimental_setup}. The chip is connected to the control setup with eight lines: the readout line, three lines for simultaneous single-qubit gates application (XY controls) and flux control (Z control), three lines connected with $10~mK$ stage and ended with $50\Omega$ terminators for qubit reset, and the coupler's control line for two-qubit gates.

Pulse generation and flux control are fully performed by a Zurich Instruments HDAWG8 arbitrary waveform generator. One analog output port of the generator is used per fluxonium circuit. IQ microwave mixers are employed to up- and downconvert the intermediate-frequency readout pulses to the resonator frequencies and back. After getting reflected from the qubit chip, the readout microwave signal is measured by a vector network analyzer (R\&S ZVB20) for spectroscopy and a home-built digitizer setup for single-shot readout. For mixer calibration we use a spectrum analyzer (Agilent N9030A).

Microwave attenuators are used to isolate the qubit chip from thermal and instrumental noise from the signal sources, which are located at room temperature. The readout line is equipped with an impedance matched parametric amplifier (IMPA) followed by a Quinstar CWJ1019KS414 isolator to prevent noise from higher-temperature stages entering the IMPA and the qubit device. We pump the IMPA using a Agilent E8257D signal generator. Three Raditec RADC-4.0-8.0-Cryo circulators and a set of low-pass and high-pass filters placed after the sample allows for the signal to pass through to the IMPA without being attenuated while removing all the reflected noise of the IMPA and dumping it into a $50~\Omega$ termination. At the PT2 stage (3 K) of the cryostat, a LNF-LNC0.3\_14A high electron mobility transistor (HEMT) is installed. The output line is further amplified outside the cryostat with two Mini-Circuits ZVA-183-S+ amplifiers. 

Due to the large DC contol current $\sim 83~\mathrm{\mu A}$ required for qubit biasing, reset pulses and control, we cannot install attenuators on the 10~mK stage of the cryostat to suppress thermal noise from room temperature microwave sources. Instead we use a low-pass filter (Mini Circuits VLF-630+) in combination with a powder filter with 15 dB attenuation close to the qubit frequencies.
Capacitively coupled qubit control lines are connected to $50~\Omega$ terminators at the 10 mK stage of the cryostat. Due to the low capacitive coupling, these lines have no effect when the qubit frequency is low, but result in enhanced decay when the bias flux is close to zero. This effect is used for qubit initialization and reset \ref{sec:Qubits_initialization}.

\section{\label{sec:Device_parameters}Device parameters}

The harmonic and fluxonium degrees of freedom of a single fluxonium circuit are described by the Hamiltonian \cite{Moskalenko2021}:

\begin{equation}
\label{eq:C1}
\hat{H} = \hat{H}_\textnormal{h} + \hat{H}_\textnormal{f}, 
\end{equation}

\begin{equation}
\label{eq:C2}
\hat{H}_\textnormal{h} =  4E_{\textnormal{Ch}}(\hat{n}^+)^2 + \frac{1}{2}E_\textnormal{L}(\hat{\vartheta}^+-{\varphi}^\textnormal{x})^2 , 
\end{equation}

\begin{equation}
\label{eq:C3}
\hat{H}_\textnormal{f} = 4E_{\textnormal{Cf}}(\hat{n}^-)^2  + \frac{1}{2}E_\textnormal{L}(\hat{\vartheta}^--{\varphi}^\textnormal{x})^2 + E_{\textnormal{J}}[1-\cos(\hat{\vartheta^-})],
\end{equation}

where $E_\textnormal{L}$, $E_{\textnormal{Cf}}$, $E_{\textnormal{J}}$ are the inductive energy, the charging energy and the Josephson junction energy of the fluxonium mode and $E_{\textnormal{Ch}}$ is the charging energy of the harmonic mode. Operators $\hat{n}^-$ and $\hat{n}^+$ are the canonically conjugate Cooper pair numbers to $\hat{\vartheta}^-$ and $\hat{\vartheta}^+$, respectively. A dimensionless variable for external flux is $\varphi^\textnormal{x}=\frac{2\pi{\Phi}^\textnormal{x}}{\Phi_0}$.

The device parameters are summarized in Table~\ref{tab:device-parameters}. $E_\mathrm{Ch}, E_\mathrm{Cf}, E_\mathrm{L}, E_{\mathrm{J}}$ are extracted by fitting the two-tone spectroscopy data. A sample two-tone spectrum for qubit S is shown in fig.~\ref{fig:Spectr_reset}a. The circuit eigenstates are labeled as $|n_\textnormal{h}, n_\textnormal{f}\rangle$, where $n_\textnormal{h}$ is the harmonic mode occupancy and $n_\textnormal{f}$ is the fluxonium mode occupancy. The harmonic mode frequencies $\omega^\textnormal{h}_{10}/2\pi$, as well as the fluxonium mode transition frequencies $\omega^\textnormal{f}_{10}/2\pi$ and $\omega^\textnormal{f}_{21}/2\pi$ at the flux degeneracy point are also listed in Table~\ref{tab:device-parameters}.

\begin{table}
	\caption{Device parameters}\label{tab:device-parameters}
	\centering
	\begin{tabular}{lllc}
		\toprule
		Parameter    & Qubit A & Qubit B  & Qubit S 		
		    \\
		\colrule
		$E_\textnormal{Ch}$ (GHz)	& 0.562 & 0.562	& 0.568				\\
		$E_\textnormal{Cf}$ (GHz)	& 0.567 & 0.567	& 0.564				\\
		$E_\textnormal{L}$ (GHz)	& 0.835 & 0.835	& 0.858				\\
		$E_\textnormal{J}$ (GHz)	& 2.052 & 2.061	& 1.989				\\
		$\omega^\textnormal{f}_{10}/2\pi$ (MHz) &688.224 & 664.763	& 750.123				\\
		$\omega^\textnormal{f}_{12}/2\pi$ (GHz) &1.832 & 1.835	& 1.803				\\
		$\omega^\textnormal{h}_{10}/2\pi$ (GHz) &1.940 & 1.940	& 2.006
		\\
		$T_1$ ($\mu$s)	& 87 & 86	& 172				\\
		$T^*_2$ ($\mu$s)	& 51 & 76	& 113				\\
		$T^{E}_2$ ($\mu$s)	& 107 & 93	& 146				\\
		$\omega_{r}/2\pi$ (GHz) &7.167 & 7.383	& 6.841				\\
		$\kappa _{r}/2\pi$ (MHz) &7.6 & 7.2	& 7.4 \\
		$\chi _{r}/2\pi$ (MHz) &0.257 & 0.272 & 0.495				
		\\
		\botrule
	\end{tabular}
	%\centering
	%\begin{tabular}
	%$^f$ - transition frequencies of the fluxonium \\
	% mode at the flux degeneracy point. \\
	%$^h$ - transition frequency of the harmonic mode.\\
	%Energy decay time ($T_1$), Ramsey decay time ($T^*_2$),\\ 
	%and spin echo decay time ($T^E_2$) measured at the \\ flux %degeneracy points.\\
    %\end{tabular}
\end{table}

\section{\label{sec:Qubits_initialization} Initialization and readout}

We probe the quantum states of Qubit A and Qubit B via the dispersive readout scheme \cite{Blais2004}. Readout resonators are driven by square-shaped $1 \mathrm{\mu s}$-long microwave pulses. For both fluxoniums we discriminate two states: the ground state $|0\rangle$, and the first excited state $|1\rangle$. 

\begin{figure}
    \includegraphics[width=1\columnwidth]{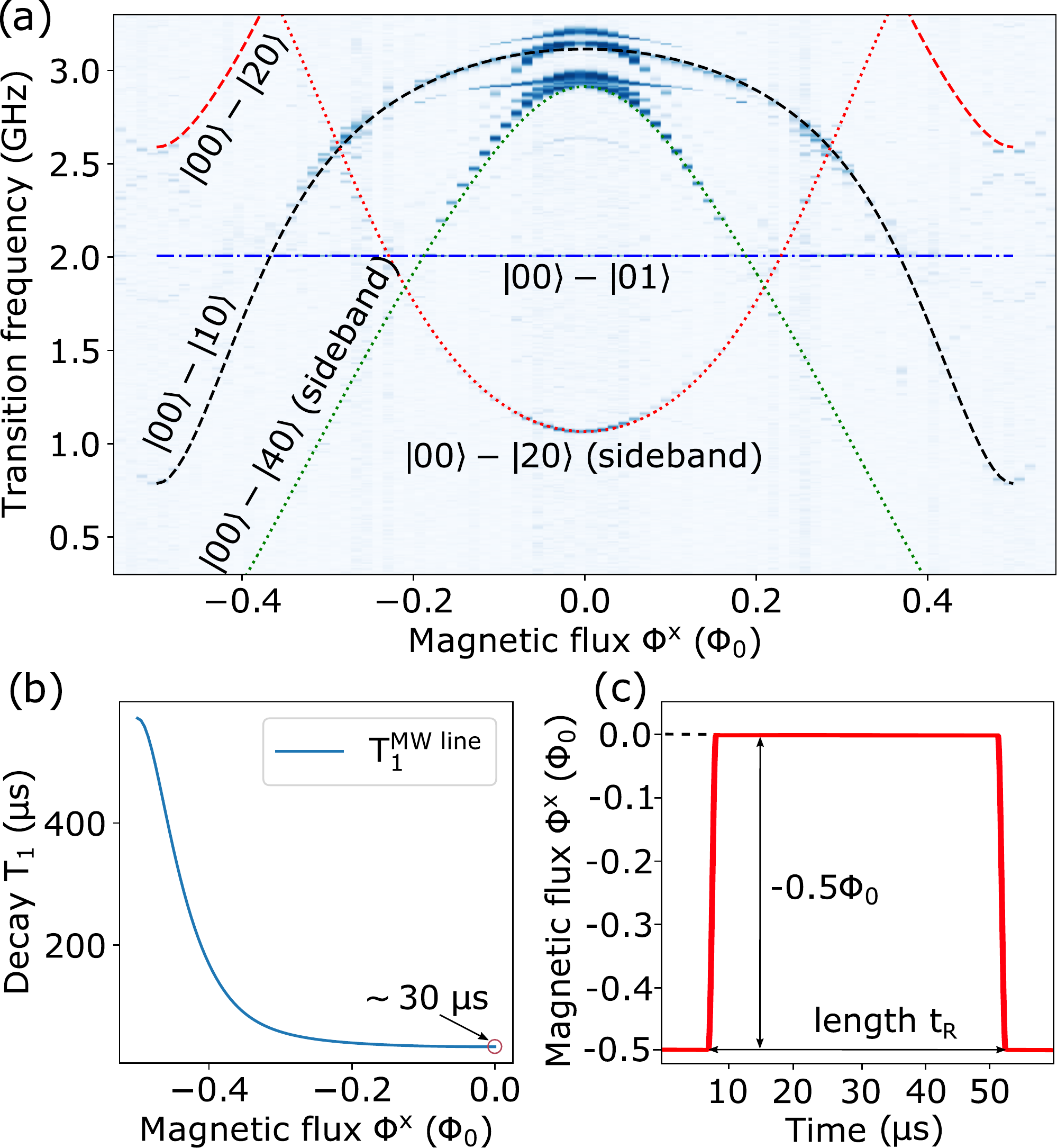}
    \caption{Spectroscopy and qubit reset protocol. (a) Two-tone response from $Q_S$ as a function of probe tone frequency and magnetic flux $\Phi^\textnormal{x}$ threading the qubit loop. The parameters of the system are extracted by fitting the spectroscopy data to the eigenenergies of the Hamiltonian eq.\eqref{eq:C1}. (b) Calculation of the energy relaxation time $T_1$ vs magnetic flux $\Phi^\textnormal{x}$ in the qubit loop through the capacitively coupled antenna. (c) Shape of the flux pulse used for qubit initialization.}
    \label{fig:Spectr_reset}
\end{figure}

Low qubit frequencies $\hbar \omega_{01} \le k_B T$ lead to operation in relatively ``hot'' environment. Unlike the case of transmon qubits, energy relaxation steers the fluxonium degree of freedom into a high entropy mixed state. Each qubit is susceptible to various noise sources, including noise from the flux control line, the capacitively coupled antenna, and dielectric loss. Each noise source can be characterized by an effective temperature, which may or may not be related to the physical temperature of the substrate. Despite using powder filters, we see a $\sim0.4$ equilibrium population for qubits A and B, and $\sim0.2$ equilibrium population for qubit S, which corresponds to an effective qubit temperature of $\sim80~\mathrm{mK}$ for qubits A and B and $\sim25~\mathrm{mK}$ for qubit S. For better signal-to-noise ratio in the calibration and characterization measurements, we use a qubit reset procedure before each measurement. 

Due to the presence of weakly coupled but high effective temperature noise sources, even higher frequency transmon qubits require such initialization if high fidelity pure states are required\cite{Magnard2018}. Similar to transmons, fluxoniums can be initialized in the ground state with sideband cooling \cite{Zhang2021, Bao2021} or an active reset procedure based on feed-forward control \cite{Gebauer2020}.

In our work, we utilize the capacitively coupled microwave antennas as an engineered dissipation channel to reduce the residual thermal excitation and to reset the qubit. The decay rate due to this dissipation process can be obtained through Fermi's Golden rule:
\begin{equation}
    \gamma = (2\pi)\omega\frac{Z_0}{R_Q}\frac{C_c^2}{(C_c+C_j)^2}|\langle0|\hat{n}_q|1\rangle|^2,
\end{equation}
where $C_c=0.34~\textnormal{fF}$ is the capacitive coupling with microwave antenna, $C_j=34.34~\textnormal{fF}$ is the effective qubit capacitance (it can be obtained from $E_\textnormal{Cf}$, see Table \ref{tab:device-parameters}), $\omega$ is the qubit frequency, $Z_0 = 50\Omega$ is the antenna impedance, $R_Q$ is the von Klitzing constant, and $\langle 0 |\hat{n}_q|1\rangle$ is the qubit charge matrix element for the fundamental transition. This energy relaxation into microwave antenna contributes to $T_1$ limitation, which is shown in fig.~\ref{fig:Spectr_reset}b for different magnetic flux values threading the qubit loop. It can be seen that minimal value of $T_1 \approx 30~\mathrm{\mu s}$ is reached at $\Phi^\textnormal{x}=0$. Our unconditional reset mechanism is based on this relaxation time $T_1 (\Phi^\textnormal{x})$ dependence. We use a fast but adiabatic flux pulse with $t_\textnormal{R}=30~\mathrm{\mu s}$ duration to move the qubit away from the flux degeneracy point to the zero flux point back before the main pulse sequence (see fig.~\ref{fig:Spectr_reset}c). During this flux pulse the target qubit is reset to the ground state. 

\begin{figure}
    \includegraphics[width=1\columnwidth]{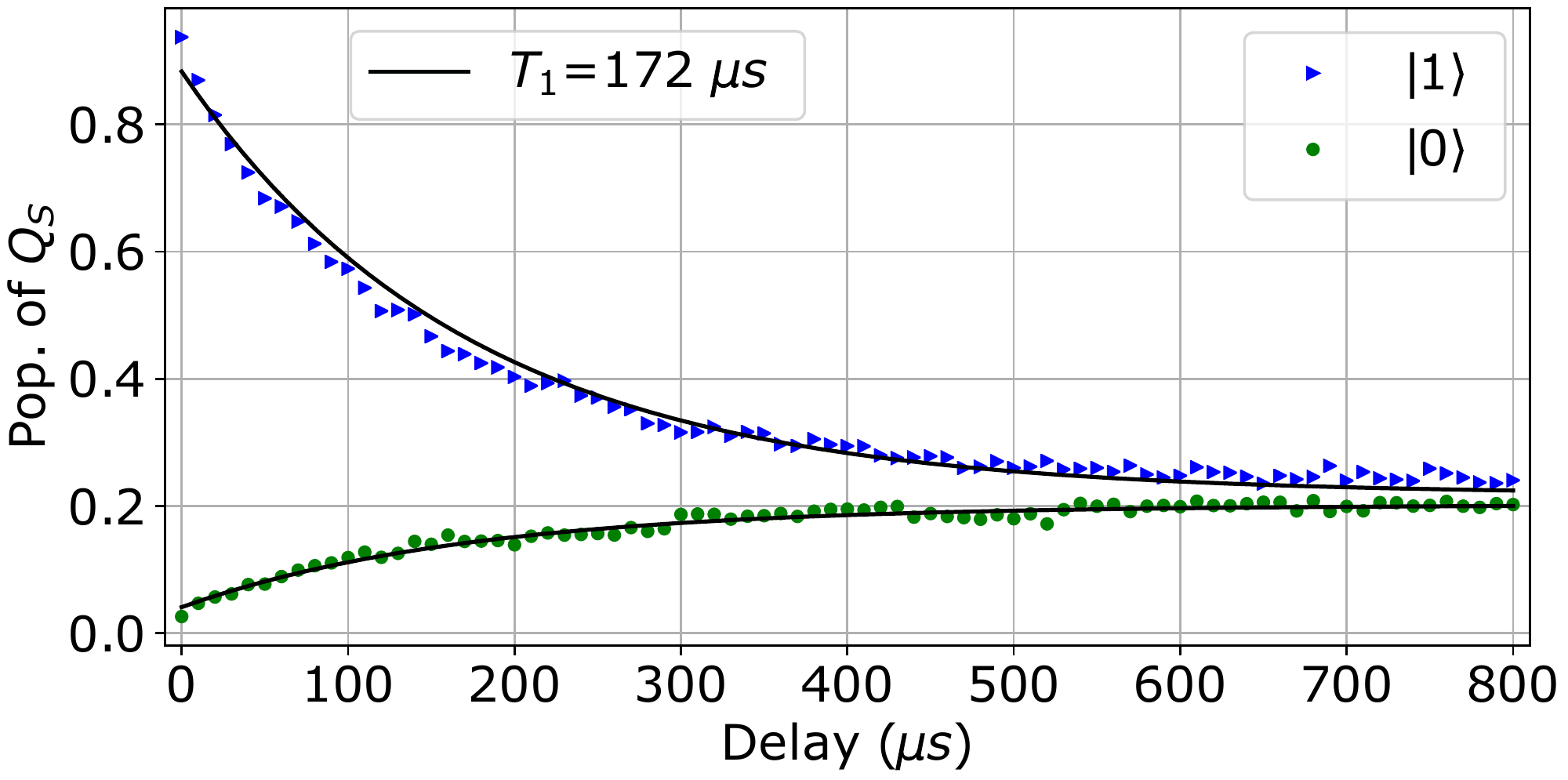}
    \caption{Time evolution of Qubit S from the ‘ground’ $|0\rangle$ (lower curve) and ‘excited’ $|1\rangle$ (upper curve) state after initialization. The solid lines are least-squares fits with exponents yielding $T_1=172 \pm 5 \mathrm{\mu s}$.}
    \label{fig:time_evolution}
\end{figure}

\begin{figure}
    \includegraphics[width=1\columnwidth]{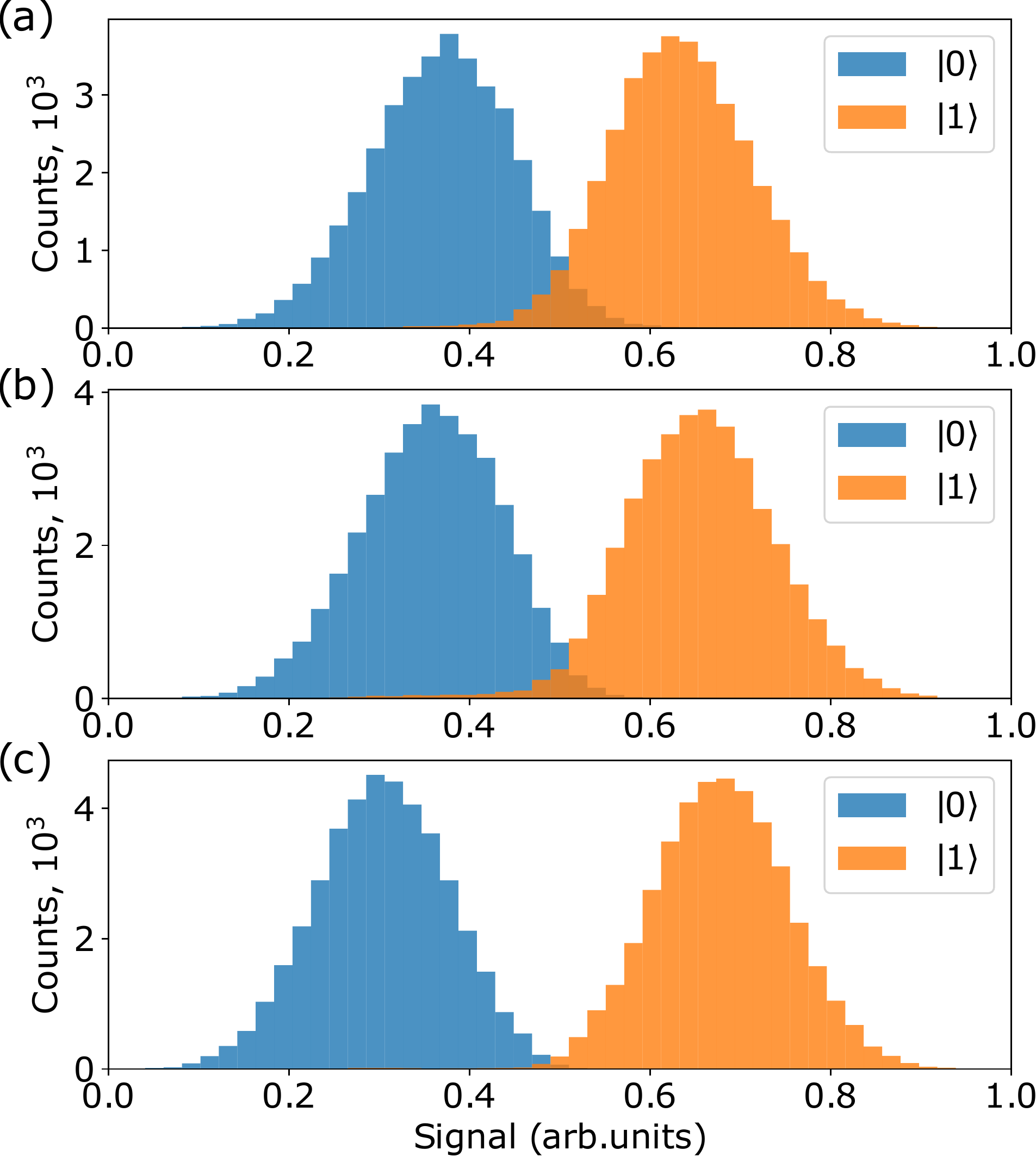}
    \caption{Histograms of the single-shot readout signals for Qubit A (a), Qubit B (b) and Qubit S (c) with preliminary qubit reset.}
    \label{fig:Single_shot}
\end{figure}

Fig.\ref{fig:time_evolution} shows the evolution of Qubit S after a reset pulse ($|0\rangle$) and after a reset pulse followed by a $\pi$ pulse ($|1\rangle$). 
Another source of noise in the measurements is readout error. We cannot directly separate the influence of initialization and readout errors. However, from the readout signal histograms shown in~Fig.~\ref{fig:Single_shot} we can conclude that the limiting factor in our measurements is separation infidelity. The readout
visibilities for Qubit A, Qubit B, and Qubit S are $87\%$, $89\%$, and $96\%$, respectively.

We attribute the relatively poor readout fidelity to deficiencies in our readout system, specifically to phase drifts in the digitizer setup and readout resonator design. Specifically, the low $\chi \ll \kappa$ ratio leads to large susceptibility to jitter between the DAC and ADC clocks.

\section{\label{sec:Parametric_drive}Parametric Z-gate calibration}

In this appendix we describe the calibration measurement procedure and results for the parametric frequency shift experiment that is used for the parametric-resonance two-qubit fSim gate described in Section~\ref{sec:TWO_qubit_gates}.

\begin{figure}
    \includegraphics[width=\columnwidth]{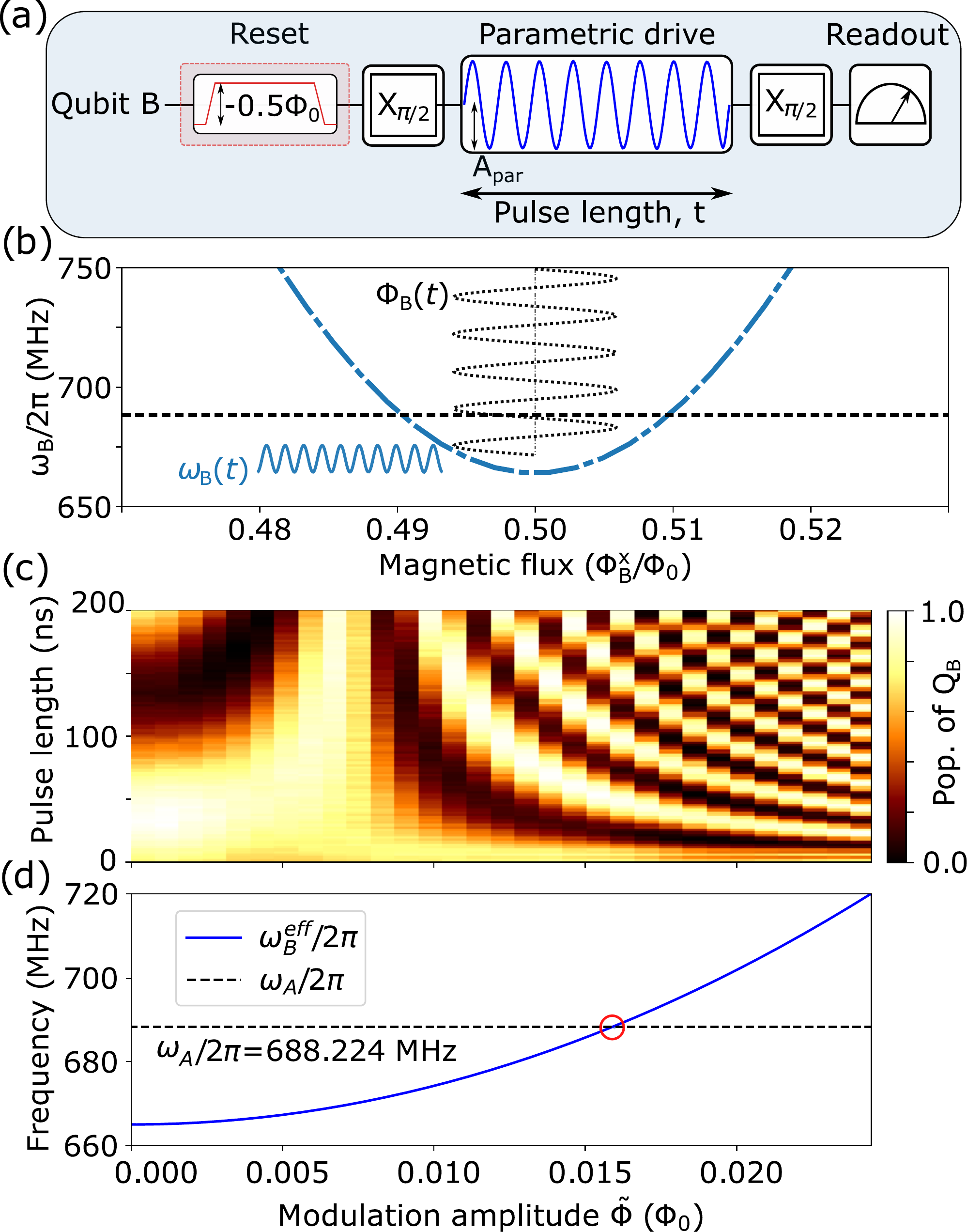}
    \caption{Parametric Z-gate calibration. (a) Pulse sequence of a Ramsey-type experiment on qubit B with an intermediate parametrically modulated Z-gate. (b) Transition frequency of the tunable qubit B (dash-dotted blue curve) and non tunable qubit A (horizontal dashed line). Biasing the flux near the flux degeneracy point ensures long coherence times by removing first order sensitivity to flux noise. The effect of the flux modulation $\Phi_B(t)$ (dotted sine curve) on qubit B frequency $\omega_B(t)$ is shown as solid curve. (c) Ramsey fringes of the qubit B as a function of the z-gate amplitude (d) Effective frequency of qubit B $\omega^{\mathrm{eff}}_B$ vs amplitude of the parametrically modulated Z-gate. Red circle marks the intersection point $\omega^{\mathrm{eff}}_B=\omega_A$.}
    \label{fig:Parametric_frequency_offset}
\end{figure}

To identify the parametric Z-gate operating point we perform Ramsey-like measurements with an intermediate parametric flux drive:
\begin{equation}
\Phi(t) = \bar{\Phi} + \tilde{\Phi}\operatorname{sin}\left(\omega_P t_d\right).
\label{F1}
\end{equation}

As shown in Fig.\ref{fig:Parametric_frequency_offset}a, this parametric flux drive for qubit B is realised by applying a harmonic signal to the individual control line of the qubit. During our experiments qubit B initially is biased at the flux degeneracy point $\bar{\Phi} = \frac{\Phi_0}{2}$ and the modulation frequency is $\omega_P/2\pi = 100~\mathrm{MHz}$. The qubit frequency dependence becomes
\begin{equation}
\omega^{\mathrm{eff}}_B(t) = \omega_B^f + \frac{\tilde{\Phi}^2}{4}\frac{\partial^2 \omega_B^f}{\partial \Phi^2}\left( 1 - \operatorname{cos}\left(2\omega_P t_d\right) \right)
\label{F3}
\end{equation}
for small $\tilde{\Phi}$. The reason for frequency doubling in eq.~\eqref{F3} is that qubit B frequency undergoes two cycles for each cycle of flux as shown in Fig.\ref{fig:Parametric_frequency_offset}b. 

In Ramsey-type measurements the population of qubit B is measured as a function of the parametric pulse duration ($t_d$) and the parametric drive amplitude $\tilde{\Phi}$ [Fig.\ref{fig:Parametric_frequency_offset}c].
From these results we calculate and plot the effective frequency of qubit B  versus the modulation amplitude $\tilde{\Phi}$ [Fig.\ref{fig:Parametric_frequency_offset}d]. The amplitude that corresponds to $\omega^{\mathrm{eff}}_B=\omega_A$ is the parametric offset gate operating point that is required for the two-qubit fSim gate $\tilde{\Phi}=0.016 \Phi_0$. %$A=38.817 mV$ 

\section{\label{sec:Coherent} Coherence measurements}

\begin{figure}
    \includegraphics[width=1\columnwidth]{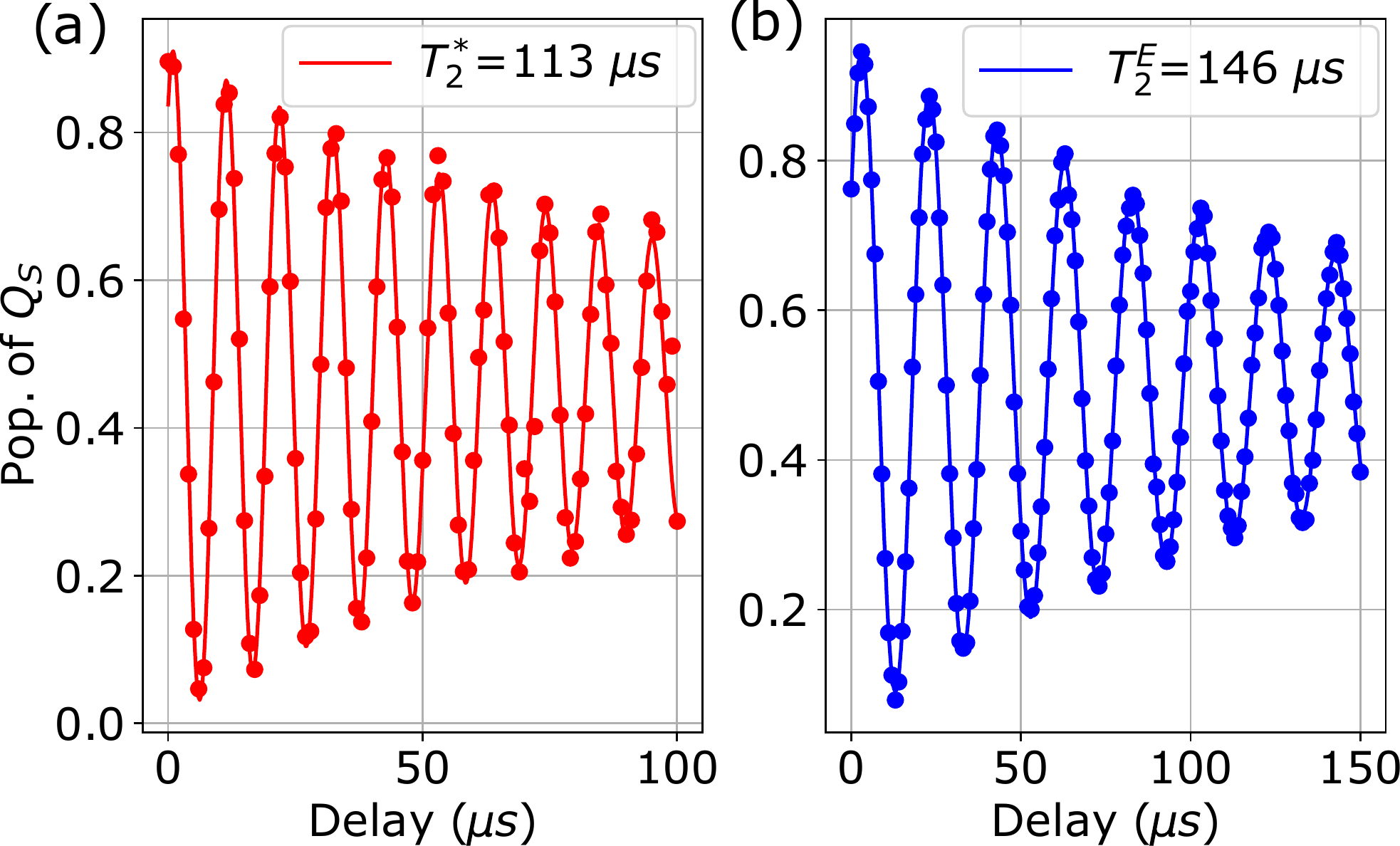}
    \caption{Coherence measurements of $Q_S$. (a) Ramsey fringes. The solid line is the fit to a decaying sinusoid with characteristic time $T^*_2=113.0 \pm 1.5 \mathrm{\mu s}$. (b) Spin echo data. The solid line is the fit to a decaying sinusoid with characteristic time $T^E_2=146.0 \pm 0.3 \mathrm{\mu s}$.}
    \label{fig:Times}
\end{figure}

In this appendix we describe the measurements of coherence times $T^*_2$ and $T^\textnormal{E}_2$ of the qubits at their flux degeneracy points [see. Table \ref{tab:device-parameters}]. The coupler is biased at zero flux. The experimental data for Qubit S is shown in Fig.\ref{fig:Times}.  $T^*_2$ is measured with a Ramsey oscillation experiment. $T^E_2$ is measured with a modified spin echo protocol: the phase of the final $\pi/2$-pulse is scanned along with the delay, giving rise to oscillations. We manually introduce these oscillations to ensure better parameter estimation from fitting. $T_1$ is obtained from the time evolution data shown in Fig.\ref{fig:time_evolution}.

\begin{figure}
    \includegraphics[width=1\columnwidth]{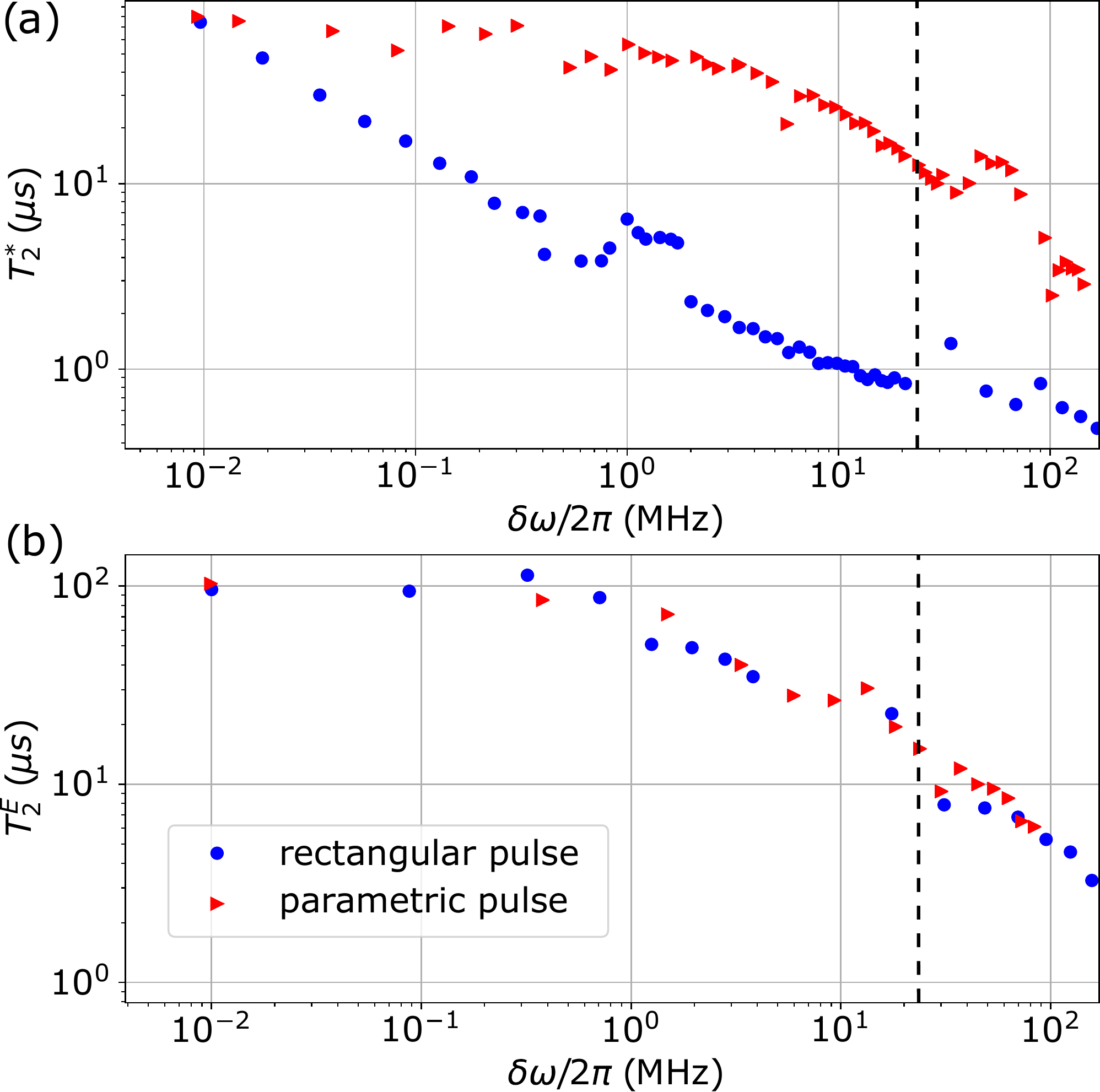}
    \caption{Coherence times $T^*_2$ and $T^\textnormal{E}_{2}$ of $Q_\textnormal{B}$ vs qubit frequency. (a) Coherence time $T^*_2$ of $Q_\textnormal{B}$ as a function of qubit offset frequency $\delta \omega / 2\pi$. (b) Coherence time $T^\textnormal{E}_{2}$ as a function of qubit offset frequency $\delta \omega / 2\pi$. Vertical dashed lines indicate the frequency offset $\delta \omega / 2\pi = 23.461~MHz$, when $\omega_B = \omega_A$.}
    \label{fig:Times_frequency_dependence}
\end{figure}

We also measure $T^*_2$ and $T^\textnormal{E}_{2}$ of Qubit B as functions of its frequency offset $\delta \omega = \omega_B - \omega_B(\Phi_0/2)$ [Fig.~\ref{fig:Times_frequency_dependence}] from the flux degeneracy point ($\omega_B(\Phi_0/2) =664.763~\mathrm{MHz}$). Here we use the measurement protocol which was described in \cite{T2_zpulse}. Between the two $\pi/2$ pulses, a rectangular or sine-shaped flux pulse with 100~MHz frequency is applied to the qubit bias line. Qubit control and readout are performed at the flux sweet spot. Vertical dashed lines indicate the frequency offset of qubit B when its frequency tunes to be equal to qubit A frequency.  

\begin{figure}
    \includegraphics[width=1\columnwidth]{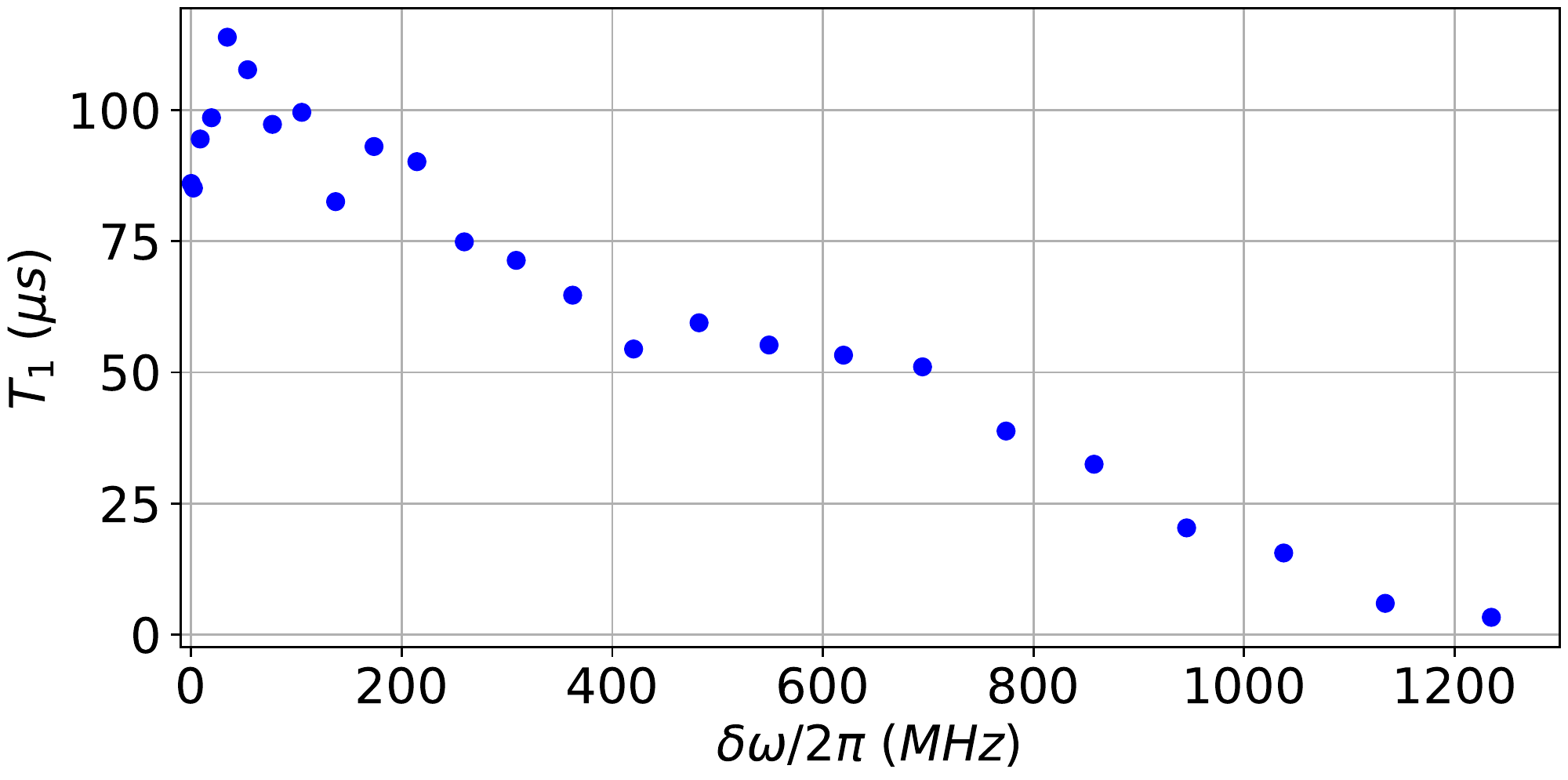}
    \caption{Measured relaxation times $T_1$ of Qubit B vs qubit frequency offset $\delta \omega/ 2\pi$.}
    \label{fig:T1_frequency_dependence}
\end{figure}

As shown in Fig.~\ref{fig:T1_frequency_dependence}, for qubit B $T_1$ varies between 80~$\mu$s and 115~$\mu$s within a 200~MHz frequency span around the flux degeneracy point.

\section{\label{sec:fSIM_calibration}Calibration of the $\sqrt{\mathrm{iSWAP}}$-like gate}

\begin{figure}
    \includegraphics[width=\columnwidth]{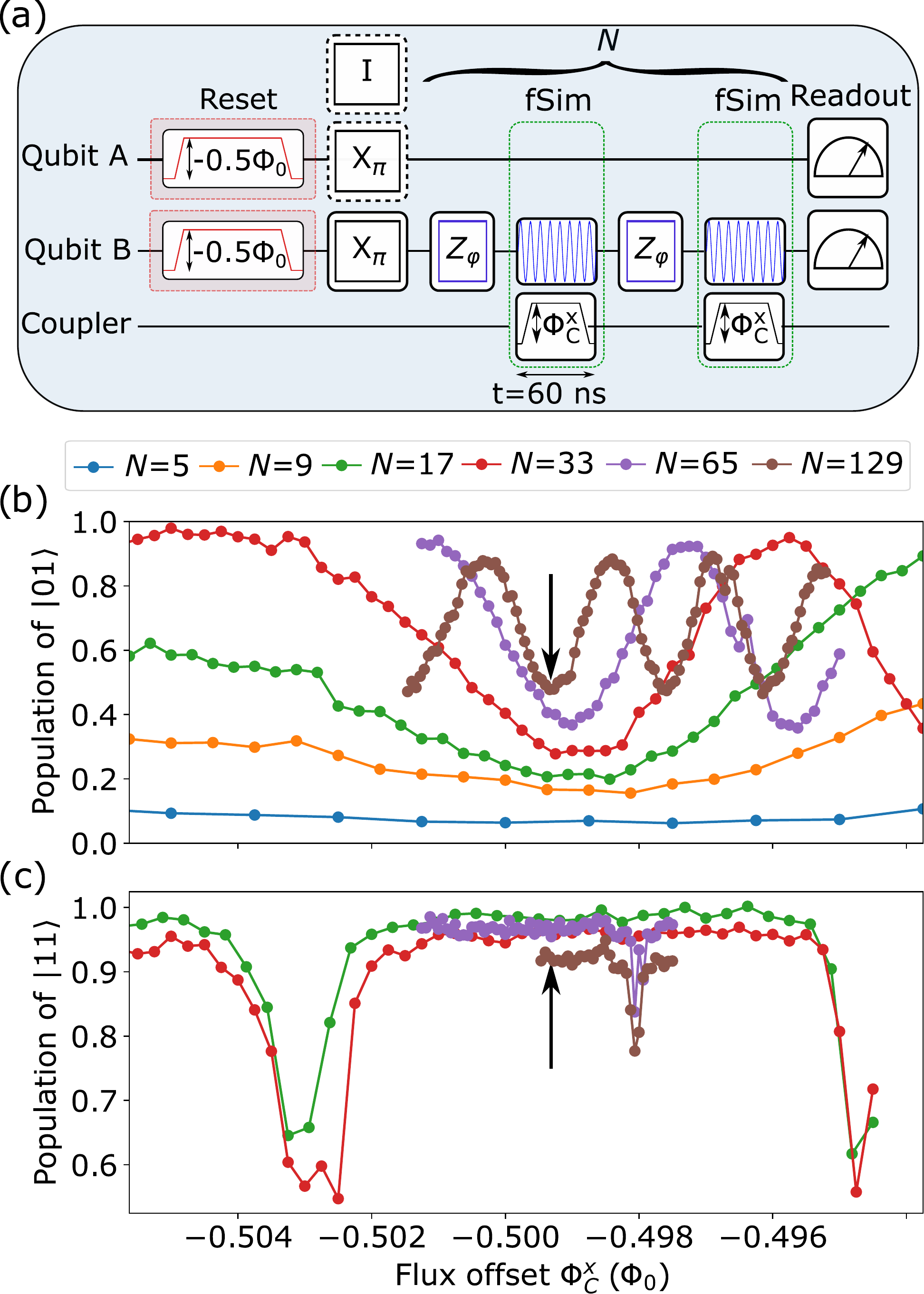}
    \caption{Calibrations for the $\sqrt{iSWAP}$-like gate. (a) Pulse sequences for the calibration of the pulse amplitude of the $\sqrt{iSWAP}$-like gate. Qubit A is prepared in the ground state ($|01\rangle$ state of the two-qubit system) for the swap amplitude calibration and in the excited state ($|11\rangle$ state of the two-qubit system) for leakage measurements. The $\sqrt{iSWAP}$-like gate consists of a parametric Z-gate applied to the Qubit B in parallel with 60-ns-long square-shaped flux pulse with 14~ns edges applied to the coupler. Parametric $Z_{\varphi}$ gates are inserted before each $\sqrt{iSWAP}$ gate for Qubit B phase compensation. (b) Measured population of $|01\rangle$ versus coupler flux offset $\Phi^x_C$. (c) Measured population of $|11\rangle$ versus coupler flux offset amplitude $\Phi^x_C$. Black arrows in (b) and (c) indicate the chosen flux amplitude $\Phi^x_C=-0.49787$.}
    \label{fig:fSIM_calibration}
\end{figure}

As described in Section~\ref{sec:iSWAP_calibration}, our two-qubit gate implementation relies on simultaneous application of a parametrically modulated Z-gate to Qubit B and a flux pulse to the coupler.
The aim of the calibration procedure is to find the amplitude $\Phi_C^x$ that corresponds to a $\sqrt{\mathrm{iSWAP}}$-like gate (fSim($\frac{\pi}{4}, \varphi$)) for a fixed gate length.
The pulse sequence used for calibration is shown in Fig.~\ref{fig:fSIM_calibration}(a). The idea behind this calibration sequence is that for an odd number $N$ of $fSim(\frac{\pi}{4}, \varphi)$ gate pairs the populations of the $|01\rangle$ and $|10\rangle$ states are swapped. In the beginning of the sequence we excite qubit B and leave qubit A unexcited. This corresponds to the $|01\rangle$ state. After the sequence, we measure the population of the $|01\rangle$ state. Our aim is to find pulse parameters that minimize the final population of $|01\rangle$ state.
Apart from the population exchange effect and the conditional phase accumulation, the two-qubit gate pulse also results in single-qubit phase shifts. If the phases accumulated by the qubits during one gate don't match, the swapping part of subsequent fSim gates will not add up coherently, and the oscillation amplitude between the states $|01\rangle$ and $|10\rangle$ will be less than unity. To avoid the effect of unequal phase accumulation, after each fSim gate we add a physical Z-gate, which is implemented as a single period of the parametric frequency shift pulse. For a correct estimate of $\Phi_C^x$ we need to simultaneously find the Z-gate amplitude that maximizes the swap amplitude.

The calibration procedure consists of sequentially scanning over the coupler flux pulse amplitude and then over the parametric Z-gate pulse amplitude. The pair of measurements is performed first for $N=5$, then for $N=9$ and further for $N=2^m+1$, where $m$ is the iteration number. 
At large $m$ decoherence becomes a limiting factor, reducing the oscillation amplitude between the $|01\rangle$ and $|10\rangle$ states. At each iteration, we update our estimate for optimal Z-gate amplitude and coupler flux pulse amplitude by choosing the minimum point of $|01\rangle$ state population. For better robustness, we add up the linearly interpolated populations from all previous iterations.

The results of the scans over $\Phi_C^x$ are shown in Fig.~\ref{fig:fSIM_calibration}(b). The calibrated value of $\Phi^x_C=-0.49787$ corresponding to $\theta=\pi/4$ is shown with a black arrow.
 
We also measure the retention of the $|11\rangle$ state. Under certain conditions, the doubly excited state may leak into a non-computational excited state of the coupler degree of freedom.
We prepare $|11\rangle$ by applying $\pi$-pulses to both qubits ($Q_A$ and $Q_B$) and measure the state population of $|11\rangle$ after a set of $\sqrt{\mathrm{iSWAP}}$ gates [Fig.~\ref{fig:fSIM_calibration}(c)]. The black arrow indicates the optimal flux amplitude.

\section{\label{sec:Crosstalk_calibration}Flux crosstalk}

In our processor, each computational qubit and coupler is controlled via individual galvanically coupled fast flux lines. The applied current not only tunes the frequency of the target qubit but may also induce additional flux threading through other qubits' loops. Here we use a Ramsey experiment with qubit flux offset to measure the DC crosstalk coefficients. 

\begin{figure}
    \includegraphics[width=\columnwidth]{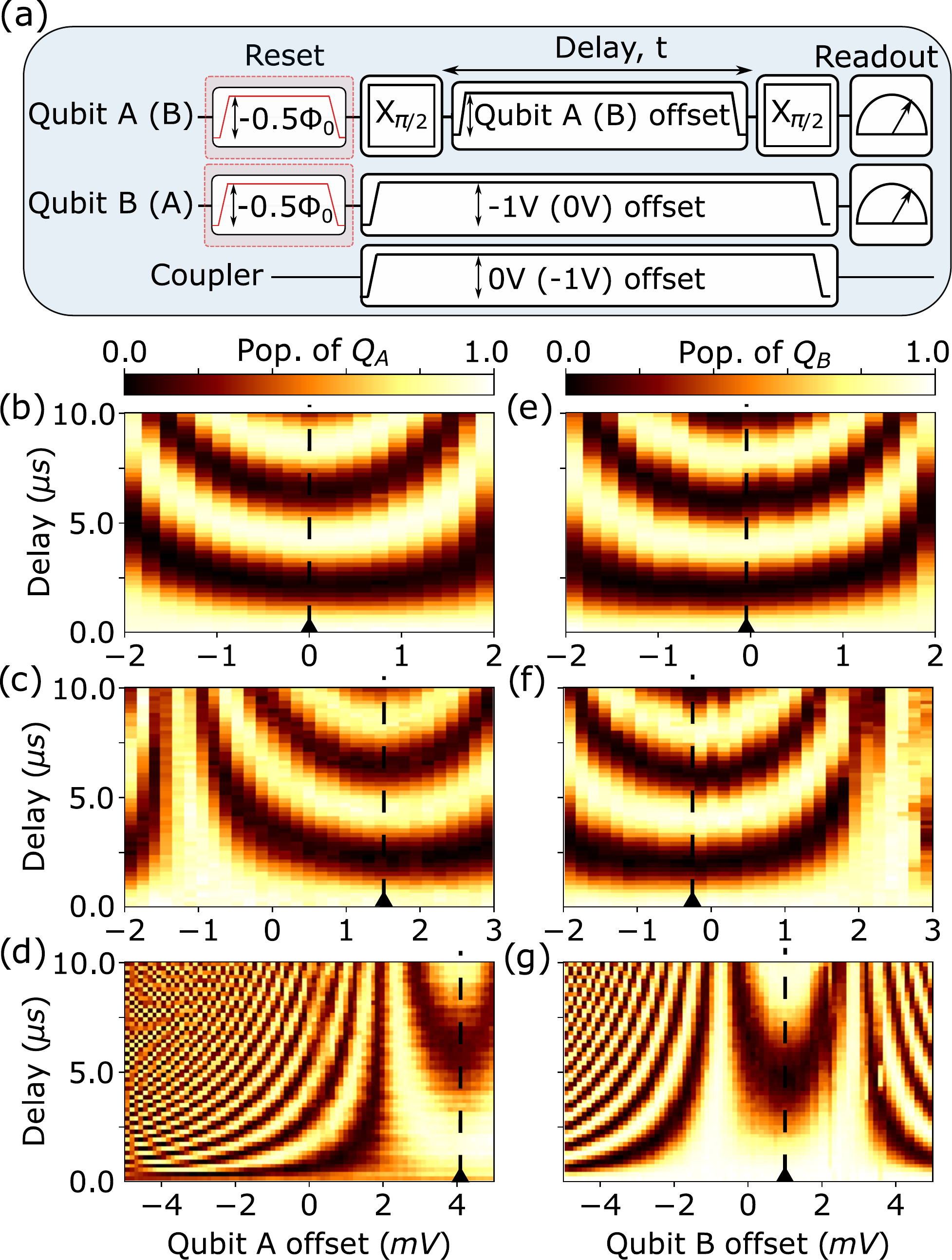}
    \caption{Flux crosstalk measurement. (a) Pulse sequences of an Ramsey-type experiment with flux offset. (b) Ramsey oscillations of the qubit A as a function of the voltage offset applied to the qubit A without flux offset on qubit B and coupler. (c) Ramsey oscillations of qubit A with -1V flux offset on qubit B. (d) Ramsey oscillations of qubit A with -1V flux offset on the coupler. (e) Ramsey oscillations of the qubit B as a function of the voltage offset applied to the qubit B without flux offset on qubit A and coupler. (f) Ramsey oscillations of qubit B with -1V offset on qubit A. (g) Ramsey oscillations of qubit B with -1V offset on the coupler.}
    \label{fig:crasstalk_measurements}
\end{figure}

We introduce a square crosstalk matrix for qubits A, B and the coupler. The diagonal elements of the crosstalk matrix are equal to unity. For the off-diagonal elements, in the $i$-th row and $j$-th column we write the equivalent signal amplitude that needs to be applied to the $i$-th control line to induce the same flux in the $i$-th fluxonium as a unit signal applied to $j$-th control line does.

The measurement of the element $(i,j)$ of the crosstalk matrix is performed by Ramsey oscillations on the $i$-th qubit, as shown in Fig.~\ref{fig:crasstalk_measurements}a. During the delay time between the two $\pi/2$ pulses of the of the Ramsey oscillation protocol, we apply a large (-1~V) signal to the $j$-th control line, and a small signal to the $i$-th control line. We sweep over the small signal applied to the $i$-th control line and the Ramsey delay time. 

In Fig.~\ref{fig:crasstalk_measurements}(c,d), we show Ramsey oscillations of qubit A as a function of an intermediate Z-pulse amplitude with $-1V$ offset on qubit B (or coupler). The similar data for qubit B with $-1V$ offset on qubit A (or coupler) are shown in Fig.~\ref{fig:crasstalk_measurements}(f,g), respectively. Flux sweet spot locations are also marked with vertical dashed lines.
Finally, from the sweet spot shifts we obtain the following crosstalk matrix:

\begin{equation}
    \left[\begin{array}{cccc}
    V_{QA}\\
    V_{QC}\\
    V_{QB} 
    \end{array}\right]=\left[\begin{array}{cccc}
    1&-0.0041&-0.0015 \\
    &1& \\
    0.0002&-0.001&1 
    \end{array}\right] \left[\begin{array}{cccc}
    V_{in,A}\\
    V_{in,C}\\
    V_{in,B} 
    \end{array}\right].
    \label{eq8}
\end{equation}

The relatively high mutual inductance~12~pH of the flux control line and the qubit loop leads to smaller control signals and ultimately smaller crosstalk. In our previous work on transmons \cite{Besedin2021, Mazhorin2022}, the design of the control line was similar to the current work, but the mutual inductance between qubit and control line was~1.6~pH, requiring $\sim$8 times more current to induce the same flux in the SQUID. The average crosstalk between qubit and the flux line of the neighboring qubit was 0.6\% (see Supplementary Material of \cite{Besedin2021} for more detail), compared to the average 0.25\% between qubits A and B and the coupler in this work.

% \nocite{*}

\bibliographystyle{unsrtnat}
\bibliography{apssamp}% Produces the bibliography via BibTeX.

\end{document}